\documentclass[article]{jss}
\usepackage{amsmath, amssymb}
\usepackage{graphicx,epsfig,subfigure} 
\usepackage{enumerate}
\usepackage{natbib}

\renewcommand{\part}[1]{\vspace{.1in}\noindent\textbf{Step #1:}}

\newcommand{\brho}{\boldsymbol{\rho}}
\newcommand{\bgamma}{\boldsymbol{\gamma}}

\newcommand{\bUpsilon} {\boldsymbol{\Upsilon}}

\newcommand{\bSigma}{\boldsymbol{\Sigma}}

\newcommand{\bbeta}{\boldsymbol{\beta}}

\newcommand{\btheta}{\boldsymbol{\theta}}

\newcommand{\bDelta}{\boldsymbol{\Delta}}
\newcommand{\bmu}{\boldsymbol{\mu}}

\newcommand{\calD}{\mathcal{D}}

\newcommand{\calG}{\mathcal{G}}

\newcommand{\calL}{\mathcal{L}}

\newcommand{\calS}{\mathcal{S}}

\newcommand{\calX}{\mathcal{X}}

\newcommand{\TBP}{\mathrm{TBP}}

\newcommand{\ICAR}{\mathrm{ICAR}}
\newcommand{\GRF}{\mathrm{GRF}}
\newcommand{\IID}{\mathrm{IID}}

\newcommand{\LDTFP}{\mathrm{LDTFP}}

\newcommand{\bE}{\mathbf{E}}

\newcommand{\bfm}{\mathbf{m}}

\newcommand{\bI}{\mathbf{I}}

\newcommand{\bfs}{\mathbf{s}}

\newcommand{\bS}{\mathbf{S}}
\newcommand{\bR}{\mathbf{R}}
\newcommand{\bv}{\mathbf{v}}
\newcommand{\bV}{\mathbf{V}}

\newcommand{\bw}{\mathbf{w}}

\newcommand{\bx}{\mathbf{x}}
\newcommand{\bX}{\mathbf{X}}

\newcommand{\bz}{\mathbf{z}}
\newcommand{\bZ}{\mathbf{Z}}
\newcommand{\bzero}{\mathbf{0}}

\newcommand{\mathR}{\mathbb{R}}

\newcommand{\ds}{\displaystyle}

\author{Haiming Zhou\\Northern Illinois University  \And 
        Timothy Hanson\\ Medtronic Inc. 
        \And Jiajia Zhang \\ University of South Carolina
    }
\title{\pkg{spBayesSurv}: Fitting Bayesian Spatial Survival Models Using \proglang{R}}

\Plainauthor{Zhou, Hanson, Zhang} 
\Plaintitle{spBayesSurv: Fitting Bayesian Spatial Survival Models Using R} 
\Shorttitle{\pkg{spBayesSurv, version 1.1.3}} 

\Abstract{
  Spatial survival analysis has received a great deal of attention over the last 20 years due to the important role that geographical information can play in predicting survival. This paper provides an introduction to a set of programs for implementing some Bayesian spatial survival models in \proglang{R} using the package \pkg{spBayesSurv}. The function \texttt{survregbayes} includes the three most commonly-used semiparametric models: proportional hazards, proportional odds, and accelerated failure time. All manner of censored survival times are simultaneously accommodated including uncensored, interval censored, current-status, left and right censored, and mixtures of these. Left-truncated data are also accommodated. Time-dependent covariates are allowed under the piecewise constant assumption. Both georeferenced and areally observed spatial locations are handled via frailties. Model fit is assessed with conditional Cox-Snell residual plots, and model choice is carried out via the log pseudo marginal likelihood, the deviance information criterion and the Watanabe-Akaike information criterion. The accelerated failure time frailty model with a covariate-dependent baseline is included in the function \texttt{frailtyGAFT}. In addition, the package also provides two marginal survival models: proportional hazards and linear dependent Dirichlet process mixture, where the spatial dependence is modeled via spatial copulas. Note that the package can also handle non-spatial data using non-spatial versions of aforementioned models. 
}
\Keywords{Bayesian nonparametric, survival analysis,  spatial dependence, semiparametric models, parametric models}
\Plainkeywords{Bayesian nonparametric, survival analysis,  spatial dependence, semiparametric models, parametric models} 

\Address{
  Haiming Zhou\\
  Division of Statistics\\
  Northern Illinois University\\
  E-mail: \email{zhouh@niu.edu}\\
  
  Timothy Hanson\\
  Strategic \& Scientific Operations\\
  Medtronic Inc., Minneapolis, Minnesota, U.S.A.\\
  E-mail: \email{tim.hanson2@medtronic.com}\\
  
  Jiajia Zhang\\
  Department of Epidemiology and Biostatistics\\
  University of South Carolina\\
  E-mail: \email{jzhang@mailbox.sc.edu}\\
}

\begin{document}


\section{Introduction}
Spatial location plays a key role in survival prediction, serving as a proxy for unmeasured regional characteristics such as socioeconomic status, access to health care, pollution, etc. Literature on the spatial analysis of survival data has flourished over the last decade, including the study of leukemia survival \citep{Henderson.etal2002}, childhood mortality \citep{Kneib2006}, asthma \citep{Li.Lin2006}, breast cancer \citep{Banerjee.Dey2005,Zhou.etal2015}, political event processes \citep{Darmofal2009}, prostate cancer \citep{Wang.etal2012,Zhou.etal2017}, pine trees \citep{Li.etal2015}, threatened frogs \citep{Zhou.etal2015b}, health and pharmaceutical firms \citep{Arbia.etal2016}, emergency service response times \citep{Taylor2017}, and many others. 

Here we introduce the \pkg{spBayesSurv} \citep{Zhou.Hanson2018} package for fitting various survival models to spatially-referenced survival data.  Note that all models included in this package can also be fit without spatial information, including nonparametric models as well as semiparametric proportional hazards (PH), proportional odds (PO), and accelerated failure time (AFT) models. The model parameters and statistical inference are carried out via self-tuning adaptive Markov chain Monte Carlo (MCMC)  methods; no manual tuning is needed. The \proglang{R} syntax is essentially the same as for existing \proglang{R} \pkg{survival} \citep{survival-package} functions.  Sensible, well-tested default priors are used throughout, however, the user can easily implement informative priors if such information is available.  The primary goal of this paper is to introduce \pkg{spBayesSurv} and provide extensive examples of its use.  Comparisons to other models and \proglang{R} packages can be found in \cite{Zhou.etal2015b}, \cite{Zhou.etal2017}, and \cite{Zhou.Hanson2017}.

Section 2 discusses \pkg{spBayesSurv}'s implementation of PH, PO, and AFT frailty models for georeferenced (e.g., latitude and longitude are recorded) and areally-referenced (e.g., county of residence recorded) spatial survival data; the functions also work very well for exchangeable or no frailties.  The models are centered at a parametric family through a novel transformed Bernstein polynomial prior and the centering family can be tested versus the Bernstein extension via Bayes factors.  All manner of censoring is accommodated as well as left-truncated data; left-truncation also allows for the inclusion of time-dependent covariates.  The LPML, DIC and WAIC statistics are available for model selection; spike-and-slab variable selection is also implemented.

In Section 3, a generalized AFT model is implemented allowing for \emph{continuous stratification}.  That is, the baseline survival function is itself a function of covariates: baseline survival changes smoothly as a function of continuous predictors; for categorical predictors the usual stratified AFT model is obtained.  Note that even for the usual stratified semiparametric AFT model with one discrete predictor (e.g., clinic) it is extremely difficult to obtain inference using frequentist approaches; see \cite{Chiou.etal2015} for a recent development.  The model fit in \pkg{spBayesSurv} actually extends discrete stratification to continuous covariates, allowing for very general models to be fit.  The generalized AFT model includes the easy computation of Bayes factors for determining which covariates affect baseline survival and whether a parametric baseline is adequate.  

Finally, Section 4 offers a spatial implementation of the completely nonparametric linear dependent Dirichlet process mixture (LDDPM) model of \cite{DeIorio.etal2009} for georeferenced data.  The LDDPM does not have one simple ``linear predictor'' as do the models in Sections 2 and 3, and therefore a marginal copula approach was taken to incorporate spatial dependence.  A piecewise-constant baseline hazard PH model is also implemented via spatial copula for comparison purposes, i.e., a Bayesian version of the model presented in \cite{Li.Lin2006}.  Section 5 concludes the paper with a discussion.

Although there are many \proglang{R} packages for implementing survival models, there are only a handful of that allow the inclusion of spatial information and these focus almost exclusively on variants of the PH model.  \proglang{BayesX} \citep{Belitz.etal2015} is an immensely powerful standalone program for fitting various generalized additive mixed models, including both georeferenced and areally-referenced frailties in the PH model.  The package \pkg{R2BayesX} \citep{Umlauf.etal2015} interfaces \proglang{BayesX} with \proglang{R}, but does not appear to include the full functionality of \proglang{BayesX}, e.g., a Bayesian approach for interval-censored data is not included. \proglang{BayesX} uses Gaussian Markov random fields for discrete spatial data.  For georeferenced frailties \proglang{BayesX} uses what have been termed ``Matern splines,'' first introduced in an applied context by \cite{Kammann.Wand2003}.  Several authors have used this approach including \cite{Kneib2006}, \cite{Hennerfeind.etal2006}, and \cite{Kneib.Fahrmeir2007}.  This approximation was termed a ``predictive process'' and given a more formal treatment by \cite{Banerjee.etal2008} and \cite{Finley.etal2009}. The \pkg{spBayesSurv} package utilizes the full-scale approximation (FSA) of \cite{Sang.Huang2012} which extends the predictive process to capture both the large and small spatial scales; see Section~\ref{sec:fsa}.  

The package \pkg{spatsurv} \citep{Taylor.Rowlingson2017} includes an implementation of PH allowing for georeferenced Gaussian process frailties.  The frailty process is approximated on a fine grid and the covariance matrix inverted via the discrete Fourier transform on block circulant matrices; see \cite{Taylor2015} for details. Taylor's approach vastly improves computation time over a fully-specified Gaussian process. The package \textbf{mgcv} \citep{Wood2017} also fits a spatial PH model by including a spatial term through various smoothers such as thin plate spline, Duchon spline and Gaussian process. All the three aforementioned R packages focus on the PH model, whereas the \pkg{spBayesSurv} includes several other spatial frailty models and two marginal copula models \citep{Zhou.etal2015b}. 

To set notation, suppose subjects are observed at $m$ distinct spatial locations $\bfs_1,\dots,\bfs_m$. Let $t_{ij}$ be a random event time associated with the $j$th subject in $\bfs_i$ and $\bx_{ij}$ be a related $p$-dimensional vector of covariates, $i=1, \ldots, m, j=1,\ldots,n_i$. Then $n=\sum_{i=1}^{m}n_i$ is the total number of subjects under consideration. Assume the survival time $t_{ij}$ lies in the interval $(a_{ij},b_{ij})$, $0\leq a_{ij}\leq b_{ij}\leq \infty$. Here left censored data are of the form $(0,b_{ij})$, right censored $(a_{ij},\infty)$, interval censored $(a_{ij},b_{ij})$ and uncensored values simply have $a_{ij}=b_{ij}$, i.e., we define $(x,x)=\{x\}$. Therefore, the observed data will be $\mathcal{D}=\{(a_{ij},b_{ij},\bx_{ij},\bfs_i); i=1,\ldots, m, j=1,\ldots,n_i\}$. For areally-observed outcomes, e.g., county-level, there is typically replication (i.e., $n_i>1$); for georeferenced data, there may or may not be replication. Note although the models are discussed for spatial survival data,  non-spatial data are also accommodated. All code below is run in \proglang{R} version 3.3.3 under the platform x86\_64-apple-darwin13.4.0 (64-bit).

\section{Semiparametric frailty models}
\subsection{Models}\label{sec:semi:frailty}
The function \texttt{survregbayes} supports three commonly-used semiparametric frailty models: AFT, PH, and PO. The AFT model has survival and density functions 
\begin{equation} \label{aft}
S_{\bx_{ij}}(t)=S_0(e^{\bx_{ij}^\top\bbeta+v_i} t), \ f_{\bx_{ij}}(t)=e^{\bx_{ij}^\top\bbeta+v_i} f_0(e^{\bx_{ij}^\top\bbeta+v_i}t), 
\end{equation}
while the PH model has survival and density functions
\begin{equation} \label{ph} 
S_{\bx_{ij}}(t)=S_0(t)^{e^{\bx_{ij}^\top\bbeta+v_i}}, \ f_{\bx_{ij}}(t)=e^{\bx_{ij}^\top\bbeta+v_i} S_0(t)^{e^{\bx_{ij}^\top\bbeta+v_i}-1} f_0(t), 
\end{equation}  
and the PO model has survival and density functions
\begin{equation} \label{po}
S_{\bx_{ij}}(t)=\frac{e^{-\bx_{ij}^\top\bbeta-v_i}S_0(t)}{1+(e^{-\bx_{ij}^\top\bbeta-v_i}-1)S_0(t)}, \ f_{\bx_{ij}}(t)=\frac{e^{-\bx_{ij}^\top\bbeta-v_i}f_0(t)}{[1+(e^{-\bx_{ij}^\top\bbeta-v_i}-1)S_0(t)]^2},
\end{equation}
where $\bbeta=(\beta_1, \ldots, \beta_p)^\top$ is a vector of regression coefficients, $v_i$ is an unobserved frailty associated with $\bfs_i$, and $S_0(t)$ is the baseline survival with density $f_0(t)$ corresponding to $\bx_{ij}=\bzero$ and $v_i=0$. Let $\Gamma(a, b)$ denote a gamma distribution with mean $a/b$ and $N_p(\bmu, \bSigma)$ a $p$-variate normal distribution with mean $\bmu$ and covariance $\bSigma$. The \texttt{survregbayes} function implements the following prior distributions: 
\begin{equation*}\label{eq:semi:priors}
\begin{aligned}
\bbeta &\sim N_p(\bbeta_0, \bS_0),\\
S_0(\cdot) |\alpha, \btheta &\sim \TBP_L(\alpha, S_{\btheta}(\cdot)), ~\alpha \sim {\Gamma} (a_0, b_0), ~\btheta\sim N_2(\btheta_0, \bV_0),\\
(v_1, \ldots, v_m)^\top|\tau &\sim \ICAR(\tau^2), ~\tau^{-2}\sim \Gamma(a_\tau, b_\tau) , \text{ or}\\
(v_1, \ldots, v_m)^\top|\tau,\phi &\sim \GRF(\tau^2, \phi), ~\tau^{-2}\sim \Gamma(a_\tau, b_\tau),  ~\phi\sim \Gamma(a_\phi, b_\phi), \text{ or}\\
(v_1, \ldots, v_m)^\top|\tau &\sim \IID(\tau^2), ~\tau^{-2}\sim \Gamma(a_\tau, b_\tau)
\end{aligned}
\end{equation*}
where $\TBP_L$, $\ICAR$, $\GRF$ and $\IID$ refer to the transformed Bernstein polynomial (TBP) \citep{Chen.etal2014, Zhou.Hanson2017} prior, intrinsic conditionally autoregressive (ICAR) \citep{Besag1974} prior, Gaussian random field (GRF) prior, and independent Gaussian (IID) prior distributions, respectively. The function argument \texttt{prior} allows users to specify these prior parameters in a list with elements defined as follows:
\begin{center}
	\centering
	\begin{tabular}{ c | c c ccccccccc} 
		\hline
		element & \texttt{maxL} & \texttt{beta0} & \texttt{S0}  & \texttt{a0} & \texttt{b0} & \texttt{theta0} & \texttt{V0} & \texttt{taua0} & \texttt{taub0} & \texttt{phia0} & \texttt{phib0}\\
		symbol & $L$ & $\bbeta_0$ & $\bS_0$ & $a_0$ & $b_0$ & $\btheta_0$ & $\bV_0$ & $a_\tau$ & $b_\tau$  & $a_\phi$ & $b_\phi$\\
		\hline
	\end{tabular}
\end{center}
We next briefly introduce these priors but leave details to \cite{Zhou.Hanson2017}. 

\subsubsection{TBP prior}
In semiparametric survival analysis, a wide variety of Bayesian nonparametric priors can be used to model $S_0(\cdot)$;
see \cite{Mueller.etal2015} and \cite{Zhou.Hanson2015} for reviews. The TBP prior is attractive in that it is centered at a given parametric family and it selects only smooth densities.  For a fixed positive integer $L$, the prior $\TBP_L(\alpha, S_{\btheta}(\cdot))$ is defined as  
\begin{equation*}\label{bp:S0}
S_0(t) = \sum_{j=1}^{L} w_{j} I(S_{\btheta}(t)|j, L-j+1), ~\bw_L \sim \textrm{Dirichlet}(\alpha, \ldots, \alpha),
\end{equation*}
where $\bw_L=(w_{1}, \ldots, w_{L})^\top$ is a vector of positive weights, $I(\cdot| a, b)$ denotes a beta cumulative distribution function (cdf) with parameters $(a,b)$, and $\{S_{\btheta}(\cdot):\btheta \in \boldsymbol{\Theta}\}$ is a parametric family of survival functions with support on positive reals $\mathR^+$. The log-logistic  $S_{\btheta}(t)=\{1+(e^{\theta_1} t)^{\exp(\theta_2)}\}^{-1}$, the log-normal $S_{\btheta}(t)=1-\Phi\{(\log t+\theta_1)\exp(\theta_2)\}$, and the Weibull $S_{\btheta}(t)=1-\exp\left\{-(e^{\theta_1} t)^{\exp(\theta_2)} \right\}$ families are implemented in \texttt{survregbayes}, where $\btheta=(\theta_1, \theta_2)^\top$. In our experience, the three centering distributions yield almost identical posterior inferences but in small samples one might be preferred. The random distribution $S_0(\cdot)$ is centered at $S_{\btheta}(\cdot)$, i.e., $E[S_0(t)|\alpha, \btheta]=S_{\btheta}(t)$. The parameter $\alpha$ controls how close the weights $\bw_j$ are to $1/L$, i.e., how close the shape of the baseline survival $S_0(\cdot)$ is relative to the prior guess $S_{\btheta}(\cdot)$. Large values of $\alpha$ indicate a strong belief that $S_0(\cdot)$ is close to $S_{\btheta}(\cdot)$; as $\alpha \rightarrow \infty$, $S_0(\cdot) \rightarrow S_{\btheta}(\cdot)$ with probability 1. Smaller values of $\alpha$ allow more pronounced deviations of $S_0(\cdot)$ from $S_{\btheta}(\cdot)$. This adaptability makes the TBP prior attractive in its flexibility, but also anchors the random $S_0(\cdot)$ firmly about $S_{\btheta}(\cdot)$: $w_j=1/L$ for $j=1,\dots,L$ implies $S_0(t)=S_{\btheta}(t)$ for $t \ge0$. Moreover, unlike the mixture of Polya trees \citep{Lavine1992} or mixture of Dirichlet process \citep{Antoniak1974} priors, the TBP prior selects smooth densities, leading to efficient posterior sampling.

\subsubsection{ICAR and IID priors}
For areal data, the ICAR prior smooths neighboring geographic-unit frailties $\bv=(v_1, \ldots, v_m)^\top$. Let $e_{ij}$ be $1$ if regions $i$ and $j$ share a common boundary and 0 otherwise; set $e_{ii}=0$. Then the $m\times m$ matrix $\bE=[e_{ij}]$ is called the adjacency matrix for the $m$ regions. The prior $\ICAR(\tau^2)$ on $\bv$ is defined through the set of the conditional distributions
\begin{equation}\label{eq:areal-prior}
v_i | \{v_j\}_{j \neq i} \sim N\left(\sum_{j=1}^m e_{ij}v_j/e_{i+}, ~\tau^2/e_{i+} \right), ~ i=1,\ldots,m,
\end{equation}
where $e_{i+} = \sum_{j=1}^m e_{ij}$ is the number of neighbors of area $\bfs_i$.  The induced prior on $\bv$ under ICAR is improper; the constraint $\sum_{j=1}^m v_j=0$ is used for identifiability \citep{Banerjee.etal2014}. Note that we assume that every region has at least one neighbor, so the proportionality constant for the improper density of $\bv$ is $(\tau^{-2})^{(m-1)/2}$ \citep{Lavine.Hodges2012}. 

For non-spatial data, we consider the independent Gaussian prior $\IID(\tau^2)$, defined as
\begin{equation}\label{eq:iid-prior}
v_1, v_2, \ldots, v_m \overset{iid}{\sim} N(0,\tau^2).
\end{equation}

\subsubsection{GRF priors}
For georeferenced data, it is commonly assumed that $v_i=v(\bfs_i)$ arises from a Gaussian random field (GRF) $\{v(\bfs), \bfs\in \calS\}$ such that $\bv=(v_1,\ldots, v_m)$ follows a multivariate Gaussian distribution as $
\bv\sim N_m(\bzero, \tau^2\bR)$, where $\tau^2$ measures the amount of spatial variation across locations and the $(i,j)$ element of $\bR$ is modeled as $\bR[i,j]=\rho(\bfs_i, \bfs_j)$. Here $\rho(\cdot, \cdot)$ is a correlation function controlling the spatial dependence of $v(\bfs)$. In \texttt{survregbayes} the powered exponential correlation function $\rho(\bfs, \bfs')=\rho(\bfs, \bfs';\phi) = \exp\{ -(\phi \|\bfs- \bfs'\|)^\nu\}$ is used, where $\phi>0$ is a range parameter controlling the spatial decay over distance, $\nu\in(0,2]$ is a pre-specified shape parameter which can be specified via \code{prior$nu}, and $\|\bfs- \bfs'\|$ refers to the distance (e.g., Euclidean, great-circle) between $\bfs$ and $\bfs'$. Therefore, the prior $\GRF(\tau^2, \phi)$ is defined as 
\begin{equation*}\label{eq:point-prior}
v_i | \{v_j\}_{j \neq i} \sim N\left(-\sum_{\{j:j\neq i\}} p_{ij}v_j/p_{ii}, ~\tau^2/p_{ii} \right), ~ i=1,\ldots,m,
\end{equation*}
where $p_{ij}$ is the $(i,j)$ element of $\bR^{-1}$.  

\subsubsection{Full-scale approximation} \label{sec:fsa}
As $m$ increases evaluating $\bR^{-1}$ from $\bR$ becomes computationally impractical. To overcome this computational issue, we consider the FSA \citep{Sang.Huang2012} due to its capability of capturing both large- and small-scale spatial dependence. Consider a fixed set of ``knots'' $\calS^* = \{\bfs^*_1, \ldots, \bfs^*_K\}$ chosen from the study region. These knots are chosen using the function \texttt{cover.design} within the \proglang{R} package \pkg{fields} \citep{Nychka.etal2015}, which computes space-filling coverage designs using the swapping algorithm \citep{Johnson.etal1990}. Let $\rho(\bfs,\bfs')$ be the correlation between locations $\bfs$ and $\bfs'$. The usual predictive process approach \citep[e.g.,][]{Banerjee.etal2008} approximates $\rho(\bfs,\bfs')$ with $ \rho_l(\bfs, \bfs') = \rho^\top(\bfs, \calS^*)\rho_{KK}^{-1}(\calS^*, \calS^*)\rho(\bfs', \calS^*)$, 
where $\rho(\bfs, \calS^*) = [\rho(\bfs, \bfs_i^*)]_{i=1}^K$ is a $K\times 1$ vector, and $\rho_{KK}(\calS^*, \calS^*)=[\rho(\bfs_i^*, \bfs_j^*)]_{i,j=1}^K$ is a $K\times K$ correlation matrix at knots $\calS^*$. However, noting that $\rho(\bfs,\bfs') = \rho_l(\bfs,\bfs') + [\rho(\bfs,\bfs')-\rho_l(\bfs,\bfs')]$,  the predictive process discards entirely the residual part $ \rho(\bfs,\bfs')-\rho_l(\bfs,\bfs')$. In contrast,  the FSA approach  approximates the correlation function $\rho(\bfs,\bfs')$ with
\begin{equation}\label{FSA}
	\rho^\dag (\bfs, \bfs') = \rho_l(\bfs, \bfs') + \rho_s(\bfs, \bfs'), 
\end{equation}
where $\rho_s(\bfs, \bfs') = \left\{ \rho(\bfs,\bfs')-\rho_l(\bfs,\bfs') \right\}\Delta(\bfs,\bfs')$ serves as a sparse approximate of the residual part. Here $\Delta(\bfs,\bfs')$ is a modulating function, which is specified so that $\rho_s(\bfs,\bfs')$ can well capture the local residual spatial dependence while still permitting efficient computation. Motivated by \cite{Konomi.etal2014}, we first partition the total input space into $B$ disjoint blocks, and then specify $\Delta(\bfs,\bfs')$ in a way such that the residuals are independent across input blocks, but the original residual dependence structure within each block is retained. Specifically, the function $\Delta(\bfs,\bfs')$ is taken to be $1$ if $\bfs$ and $\bfs'$ belong to the same block and $0$ otherwise. The approximated correlation function $\rho^\dag (\bfs, \bfs')$ in Equation~\ref{FSA} provides an exact recovery of the true correlation within each block, and the approximation errors are $\rho(\bfs,\bfs')-\rho_l(\bfs,\bfs')$ for locations $\bfs$ and $\bfs'$ in different blocks. Those errors are expected to be small for most entries because most of these location pairs are farther apart. To determine the blocks, we first use the \proglang{R} function \texttt{cover.design} to choose $B\leq m$ locations among the $m$ locations forming $B$ blocks, then assign each $\bfs_i$ to the block that is closest to $\bfs_i$. Here $B$ does not need to be equal to $K$. When $B=1$, no approximation is applied to the correlation $\rho$. When $B=m$, it reduces to the approach of \cite{Finley.etal2009}, so the local residual spatial dependence may not be well captured. 

Applying the above FSA approach to approximate the correlation function $\rho(\bfs,\bfs')$, we can approximate the correlation matrix $\bR$ with
\begin{equation}\label{FSA:rho}
\begin{aligned}
\brho_{mm}^\dag &= \brho_l + \brho_s = \brho_{mK}\brho_{KK}^{-1}\brho_{mK}^\top + \left(\brho_{mm} - \brho_{mK}\brho_{KK}^{-1}\brho_{mK}^\top \right) \circ \bDelta,
\end{aligned}
\end{equation}
where $\brho_{mK} = [\rho(\bfs_i, \bfs_j^*)]_{i=1:m,j=1:K}$, $\brho_{KK}=[\rho(\bfs_i^*, \bfs_j^*)]_{i,j=1}^K$, and $\bDelta=[\Delta(\bfs_i, \bfs_j)]_{i,j=1}^m$. Here, the notation ``$\circ$'' represents the element-wise matrix multiplication. To avoid numerical instability, we add a small nugget effect $\epsilon=10^{-10}$ when defining $\bR$, that is, $\bR=(1-\epsilon)\brho_{mm}+\epsilon \bI_m$.  It follows from Equation~\ref{FSA:rho} that $\bR$ can be approximated by
$$\bR^\dag = (1-\epsilon)\brho_{mm}^\dag+\epsilon\bI_m = (1-\epsilon)\brho_{mK}\brho_{KK}^{-1}\brho_{mK}^\top  + \bR_s,$$
where $\bR_s=(1-\epsilon)\left(\brho_{mm} - \brho_{mK}\brho_{KK}^{-1}\brho_{mK}^\top \right) \circ \bDelta + \epsilon\bI_m$. Applying the Sherman-Woodbury-Morrison formula for inverse matrices, we can approximate $\bR^{-1}$ by 
\begin{equation}\label{FSA:inv}
\begin{aligned}
\left(\bR^\dag\right)^{-1}=\bR_s^{-1} - (1-\epsilon)\bR_s^{-1}\brho_{mK} \left[\brho_{KK}+(1-\epsilon)\brho_{mK}^\top\bR_s^{-1}\brho_{mK}\right]^{-1} \brho_{mK}^\top \bR_s^{-1}.
\normalsize
\end{aligned}
\end{equation}
In addition, the determinant of $\bR$ can be approximated by 
\begin{equation}\label{FSA:det}
\det\left(\bR^\dag\right) = \det\left\{\brho_{KK}+(1-\epsilon)\brho_{mK}^\top\bR_s^{-1}\brho_{mK}\right\} \det(\brho_{KK})^{-1}\det(\bR_s).
\end{equation}
Since the $m\times m$ matrix $\bR_s$ is a block matrix, the right-hand sides of Equations~\ref{FSA:inv} and \ref{FSA:det} involve only inverses and determinants of $K\times K$ low-rank matrices and $m\times m$ block diagonal matrices. Thus the computational complexity can be greatly reduced relative to the expensive computational cost of using original correlation function for large value of $m$. However, for small $m$, e.g., $m<300$, the FSA is usually slower than direct inverse of $\bR$ due to the complexity of FSA's implementation. Note that $K$ and $B$ can be specified via \code{prior$K} and \code{prior$B}, respectively.

\subsection{MCMC}\label{sec:semi:mcmc}
The likelihood function for $(\bw_L, \btheta, \bbeta, \bv)$ is given by 
\begin{equation} \label{like} 
\calL(\bw_L, \btheta, \bbeta, \bv)=\prod_{i=1}^m \prod_{j=1}^{n_i} \left[S_{\bx_{ij}}(a_{ij})-S_{\bx_{ij}}(b_{ij})\right]^{I\{a_{ij}<b_{ij}\}} f_{\bx_{ij}}(a_{ij})^{I\{a_{ij}=b_{ij}\}}.
\end{equation}
MCMC is carried out through an empirical Bayes approach \citep{Carlin.Louis2010} coupled with adaptive Metropolis samplers \citep{Haario.etal2001}.  Recall that $w_{j}=1/L$ implies the underlying parametric model with $S_0(t)=S_{\btheta}(t)$.  Thus, the parametric model provides good starting values for the TBP survival model. Let $\hat{\btheta}$ and $\hat{\bbeta}$ denote the parametric estimates of $\btheta$ and $\bbeta$, e.g., maximum likelihood estimates, and let $\hat{\bV}$ and $\hat{\bS}$ denote their estimated covariance matrices, respectively. Set $\bz_{L-1}=(z_1,\ldots,z_{L-1})^\top$ with $z_j=\log(w_{j})-\log(w_{L})$. The $\bbeta$, $\btheta$, $\bz_{L-1}$, $\alpha$ and $\phi$ are all updated using adaptive Metropolis samplers, where the initial proposal variance is $\hat{\bS}$ for $\bbeta$, $\hat{\bV}$ for $\btheta$, $0.16\bI_{L-1}$ for $\bz_{L-1}$ and $0.16$ for $\alpha$ and $\phi$. Each frailty term $v_i$ is updated via Metropolis-Hastings, with proposal variance as the conditional prior variance of $v_i | \{v_j\}_{j \neq i}$; $\tau^{-2}$ is updated via a Gibbs step from its full conditional. A complete description and derivation of the updating steps are available in \cite{Zhou.Hanson2017}. 

The function \texttt{survregbayes} sets the following hyperparameters as defaults: $\bbeta_0=\bzero$, $\bS_0=10^{10}\bI_p$, $\btheta_0=\hat{\btheta}$, $\bV_0=10\hat{\bV}$, $a_0=b_0=1$, and $a_\tau=b_\tau=.001$. Although the default $\Gamma(0.001, 0.001)$ prior on $\tau^2$ has been tested to perform well across various simulation scenarios \citep{Zhou.Hanson2017}, it still should be used with caution in practice; see \cite{Gelman2006} for general suggestions. In addition, we assume a somewhat informative prior on $\btheta$ to obviate confounding between $\btheta$ and $\bw_L$. For the GRF prior, we set $a_\phi=2$ and $b_{\phi}=(a_{\phi}-1)/\phi_0$ so that the prior of $\phi$ has mode at ${\phi}_0$ and the prior mean of $1/\phi$ is $1/\phi_0$ with infinite variance. Here $\phi_0$ satisfies $\rho(\bfs',\bfs''; \phi_0)=0.001$, where $\|\bfs'-\bfs''\|=\max_{ij}\|\bfs_i-\bfs_j\|$. Note that \cite{Kneib.Fahrmeir2007} simply fix $\phi$ at $\phi_0$, while we allow $\phi$ to be random around $\phi_0$.

\subsection{Model diagnostics and comparison}\label{sec:semi:diagnoistics}
For model diagnostics, we consider a general residual of \cite{Cox.Snell1968}, defined as $r(t_{ij}) = - \log S_{\bx_{ij}}(t_{ij}) $. Given $S_{\bx_{ij}}(\cdot)$, $r(t_{ij})$ has a standard exponential distribution. If the model is ``correct,'' and under the arbitrary censoring, the pairs $\{ r(a_{ij}), r(b_{ij}) \}$ are approximately a random arbitrarily censored sample from an $\textrm{Exp}(1)$ distribution, and the estimated \citep{Turnbull1974} integrated hazard plot should be approximately straight with slope 1. Uncertainty in the plot is assessed through several cumulative hazards based on a random posterior sample from $[\bbeta,\btheta,\bw_L,\bv|\calD]$. Note that conditional on frailties, the Cox-Snell residuals considered here are still independent.  This is in contrast to typical Cox-Snell plots which only use point estimates yieding dependent residuals under frailty models. 

For model comparison, we consider three popular model choice criteria: the deviance information criterion (DIC) \citep{Spiegelhalter.etal2002}, the log pseudo marginal likelihood (LPML) \citep{Geisser.Eddy1979}, and the Watanabe-Akaike information criterion (WAIC) \cite{Watanabe2010},  where DIC (smaller is better) places emphasis on the relative quality of model fitting, and LPML (larger is better) and WAIC (smaller is better) focus on the predictive performance. These criteria are readily computed from the MCMC output; see \cite{Zhou.Hanson2017} for more details.

\subsection{Leukemia survival data}\label{sec:semi:leukemia}
A dataset on the survival of acute myeloid leukemia in $n=1,043$ patients \citep{Henderson.etal2002} is considered, named as \code{LeukSurv} in the package. It is of interest to investigate possible spatial variation in survival after accounting for known subject-specific prognostic factors, which include \code{age}, \code{sex}, white blood cell count (\code{wbc}) at diagnosis, and the Townsend score (\code{tpi}) for which higher values indicates less affluent areas. Both exact residential locations of all patients and their administrative districts (the boundary file is named as \code{nwengland.bnd} in the package) are available, so we can fit both geostatistical and areal models. 

\subsubsection{PO model with ICAR frailties}
If the IID or ICAR frailties are considered, to easily identify the correspondence between frailties and clusters/regions, we program the function \texttt{survregbayes} so that the input dataset should be sorted by the cluster variable before any use. The following code is used to sort the dataset by \code{district} and obtain the adjacency matrix $\bE$.
\begin{CodeChunk}
\begin{CodeInput}
R> library("coda")
R> library("survival")
R> library("spBayesSurv")
R> library("fields")
R> library("BayesX")
R> library("R2BayesX")
R> data("LeukSurv")
R> d <- LeukSurv[order(LeukSurv$district), ]
R> head(d)
\end{CodeInput}
\begin{CodeOutput}
    time cens    xcoord    ycoord age sex   wbc  tpi district
24     1    1 0.4123484 0.4233738  44   1 281.0 4.87        1
62     3    1 0.3925028 0.4531422  72   1   0.0 7.10        1
68     4    1 0.4167585 0.4520397  68   0   0.0 5.12        1
128    9    1 0.4244763 0.4123484  61   1   0.0 2.90        1
129    9    1 0.4145535 0.4520397  26   1   0.0 6.72        1
163   15    1 0.4013230 0.4785006  67   1  27.9 1.50        1
\end{CodeOutput}
\begin{CodeInput}
R> nwengland <- read.bnd(system.file("otherdata/nwengland.bnd", 
+    package = "spBayesSurv"))
R> adj.mat <- bnd2gra(nwengland)
R> E <- diag(diag(adj.mat)) - as.matrix(adj.mat)
\end{CodeInput}
\end{CodeChunk}

The following code is used to fit the PO model with ICAR frailties using the TBP prior with $L=15$ and default settings for other priors. A burn-in period of 5,000 iterates was considered and the Markov chain was subsampled every 5 iterates to get a final chain size of 2,000. The argument \code{ndisplay = 1000}  will display the number of saved scans after every 1,000 saved iterates. If the argument \code{InitParamMCMC = TRUE} (not used here as it is the default setting), then an initial chain with \code{nburn = 5000}, \code{nsave = 5000}, \code{nkip = 0} and \code{ndisplay = 1000} will be run under parametric models; otherwise, the initial values are obtained from fitting parametric non-frailty models via \texttt{survreg}. The total running time is 166 seconds. 
\begin{CodeChunk}
\begin{CodeInput}
R> set.seed(1)
R> mcmc <- list(nburn = 5000, nsave = 2000, nskip = 4, ndisplay = 1000)
R> prior <- list(maxL = 15)
R> ptm <- proc.time()
R> res1 <- survregbayes(formula = Surv(time, cens) ~ age + sex + wbc + tpi +
+    frailtyprior("car", district), data = d, survmodel = "PO",
+    dist = "loglogistic", mcmc = mcmc, prior = prior, Proximity = E)
R> proc.time() - ptm
\end{CodeInput}
\begin{CodeOutput}
   user  system elapsed 
165.919   0.296 166.354
\end{CodeOutput}
\end{CodeChunk}
The term \code{frailtyprior("car", district)} indicates that the ICAR prior in Equation~\ref{eq:areal-prior} is used. One can also incorporate the IID prior in Equation~\ref{eq:iid-prior} via \code{frailtyprior("iid", district)}. The non-frailty model can be fit by removing the \code{frailtyprior} term. The argument \code{survmodel} is used to indicate which model will be fit; choices include \code{"PH"}, \code{"PO"}, and \code{"AFT"}. The argument \code{dist} is used to specify the distribution family of $S_{\btheta}(\cdot)$ defined in Section~\ref{sec:semi:frailty}, and the choices include \code{"loglogistic"}, \code{"lognormal"}, and \code{"weibull"}. The argument \code{prior} is used to specify user-defined hyperparameters, e.g., for $p=3$, $L=15$, $\bbeta_0=\bzero$, $\bS_0=10\bI_p$, $\btheta_0=\bzero$, $\bV_0=10\bI_2$, $a_0=b_0=1$, and $a_\tau=b_\tau=1$, the prior can be specified as below.
\begin{CodeInput}
R> prior <- list(maxL = 15, beta0 = rep(0, 3), S0 = diag(10, 3), 
+    theta0 = rep(0, 2), V0 = diag(10, 2), a0 = 1, b0 = 1, 
+    taua0 = 1, taub0 = 1)
\end{CodeInput}
If \code{prior = NULL}, then the default hyperparameters given in Section~\ref{sec:semi:mcmc} would be used. Note by default \code{survregbayes} standardizes each covariate by subtracting the sample mean and dividing the sample standard deviation. Therefore, the user-specified hyperparameters should be based on the model with scaled covariates unless the argument \code{scale.designX = FALSE} is added. 

The output from applying the \code{summary} function to the returned object \code{res1} is given below. 
\begin{CodeChunk}
\begin{CodeInput}
R> (sfit1 <- summary(res1))
\end{CodeInput}
\begin{CodeOutput}
Proportional Odds model:
Call:
survregbayes(formula = Surv(time, cens) ~ age + sex + wbc + tpi + 
    frailtyprior("car", district), data = d, survmodel = "PO", 
    dist = "loglogistic", mcmc = mcmc, prior = prior, Proximity = E)

Posterior inference of regression coefficients
(Adaptive M-H acceptance rate: 0.2731):
      Mean        Median      Std. Dev.   95%CI-Low   95%CI-Upp 
age   0.0519835   0.0518955   0.0034329   0.0455544   0.0589767
sex   0.1238558   0.1241657   0.1061961  -0.0854203   0.3274537
wbc   0.0059439   0.0059223   0.0008163   0.0043996   0.0074789
tpi   0.0598826   0.0597254   0.0159244   0.0286519   0.0904957

Posterior inference of conditional CAR frailty variance
          Mean      Median    Std. Dev.  95%CI-Low  95%CI-Upp
variance  0.080346  0.056350  0.082950   0.001709   0.299395 

Log pseudo marginal likelihood: LPML=-5925.194
Deviance Information Criterion: DIC=11849.82
Watanabe-Akaike information criterion: WAIC=11850.39
Number of subjects: n=1043
\end{CodeOutput}
\end{CodeChunk}
We can see that \code{age}, \code{wbc} and \code{tpi} are significant risk factors for leukemia survival. For example, lower \code{age} decreases the odds of a patient dying by any time; holding other predictors constant, a 10-year decrease in age cuts the odds of dying by $\exp(-10\times 0.05)\approx 60\%$. The posterior mean for $\tau^2$ is $0.08$. The LPML, DIC and WAIC are -5925, 11850 and 11850, respectively. 

\begin{figure}
	\centering
	{ \includegraphics[width = .8\textwidth]{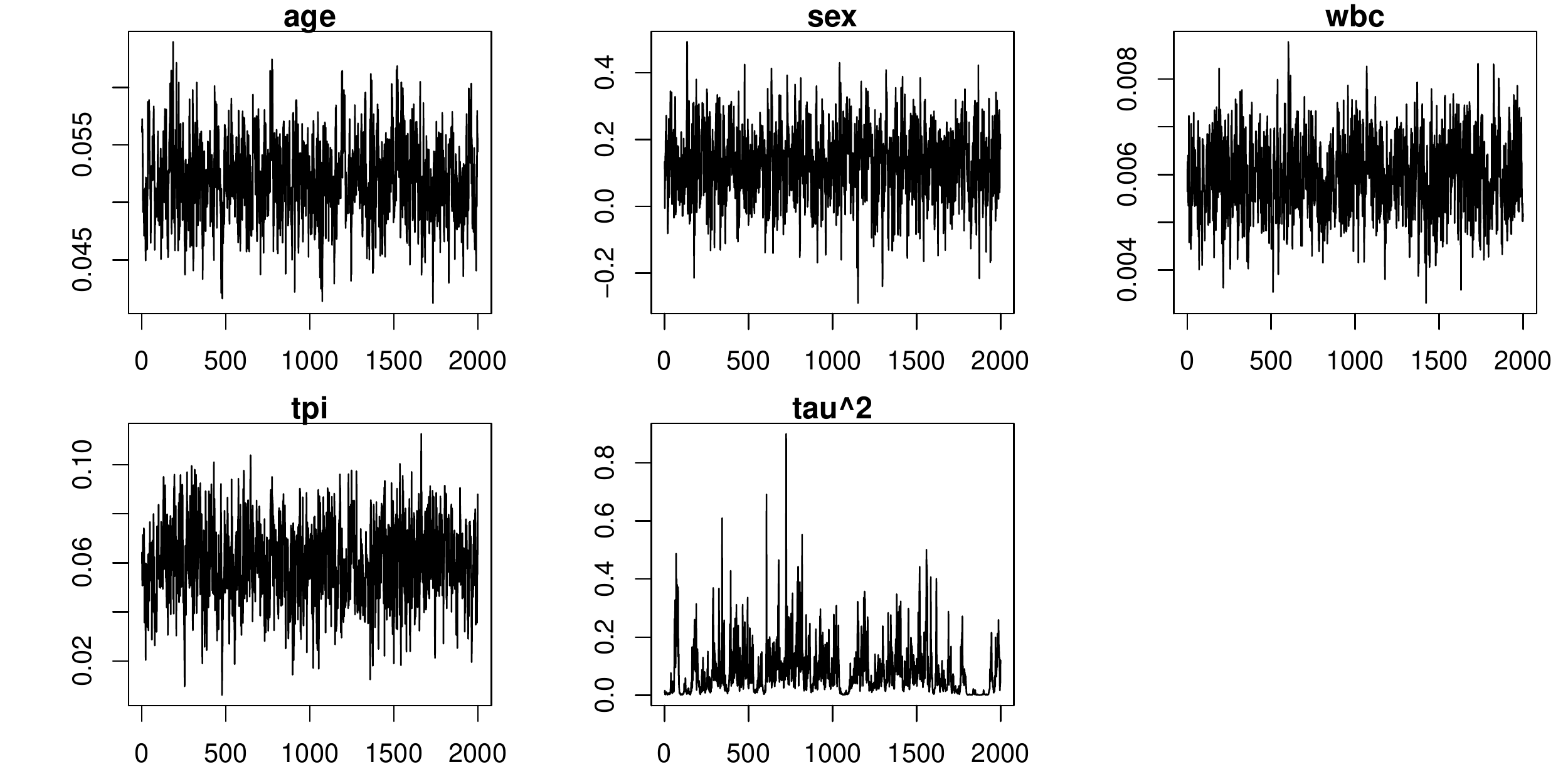} }
	\caption{Leukemia survival data. Trace plots for $\bbeta$, $\tau^2$ and $\alpha$ under the PO model with ICAR frailties.}
	\label{leukemia:trace}
\end{figure}

The following code is used to produce trace plots (Figure~\ref{leukemia:trace}) for $\bbeta$ and $\tau^2$. Note that the mixing for $\tau^2$ is not very satisfactory. This is not surprising, since we are using very vague gamma prior $\Gamma(0.001, 0.001)$ and the total number of districts is only $24$. One may consider to use a more informative prior $\Gamma(1, 1)$ on $\tau^2$ or run a longer chain with higher thin to improve the mixing. 
\begin{CodeInput}
R> par(mfrow = c(3, 2)) 
R> par(cex = 1, mar = c(2.5, 4.1, 1, 1))
R> traceplot(mcmc(res1$beta[1,]), xlab = "", main = "age")
R> traceplot(mcmc(res1$beta[2,]), xlab = "", main = "sex")
R> traceplot(mcmc(res1$beta[3,]), xlab = "", main = "wbc")
R> traceplot(mcmc(res1$beta[4,]), xlab = "", main = "tpi")
R> traceplot(mcmc(res1$tau2), xlab = "", main = "tau^2")
\end{CodeInput}

\begin{figure}
	\centering
	\subfigure[]{ \includegraphics[width=0.3\textwidth]{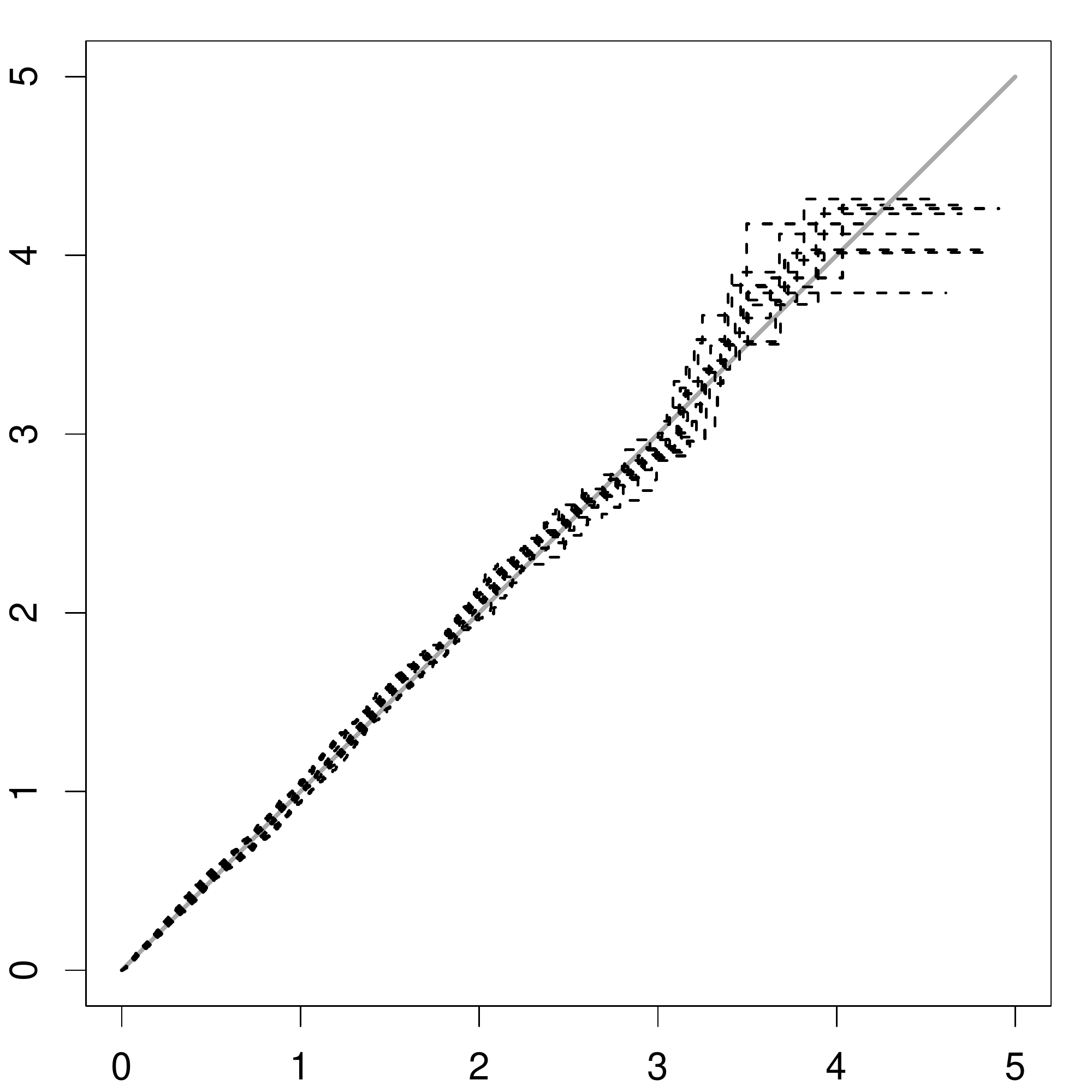} }
	\subfigure[]{ \includegraphics[width=0.3\textwidth]{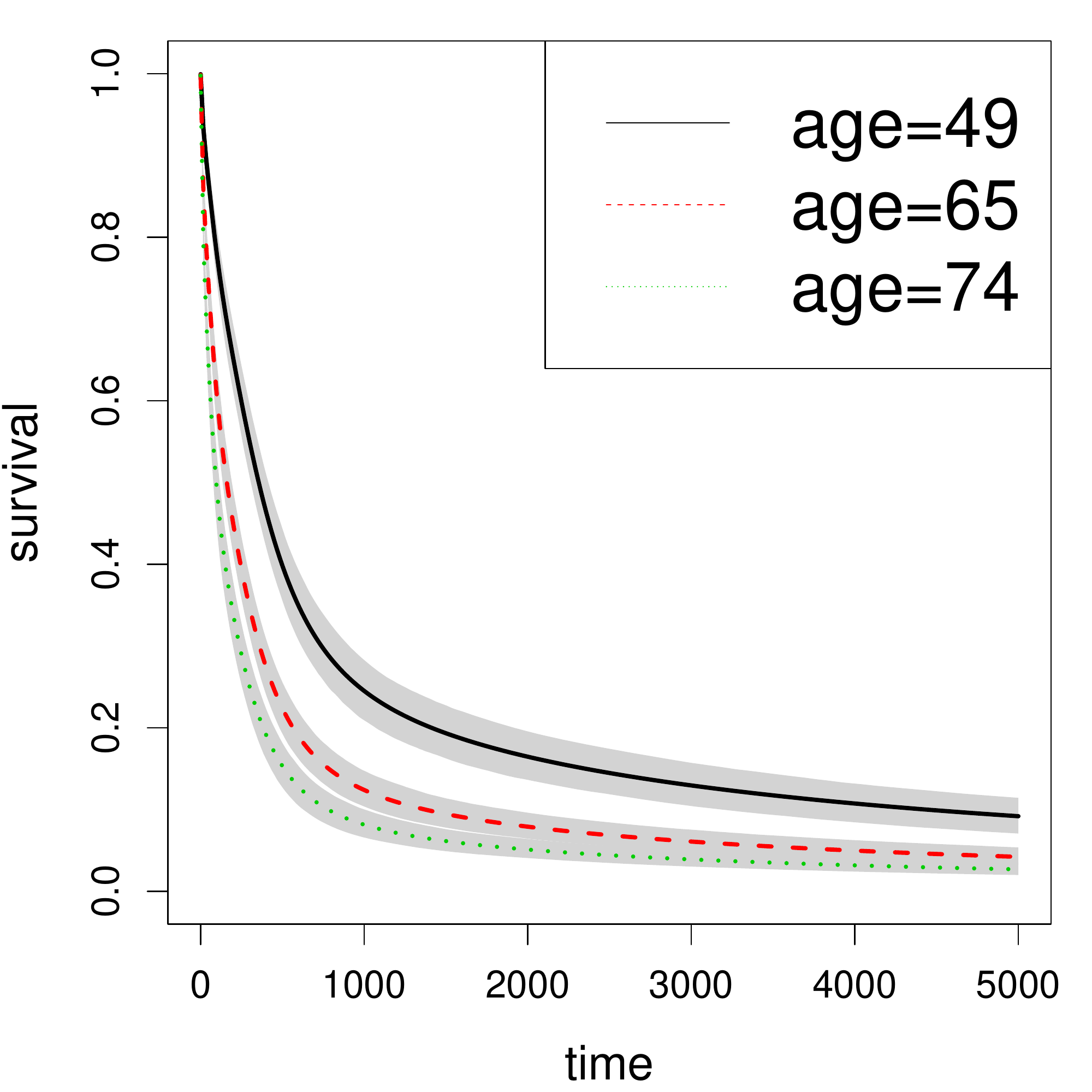} }
	\subfigure[]{ \includegraphics[width=0.3\textwidth]{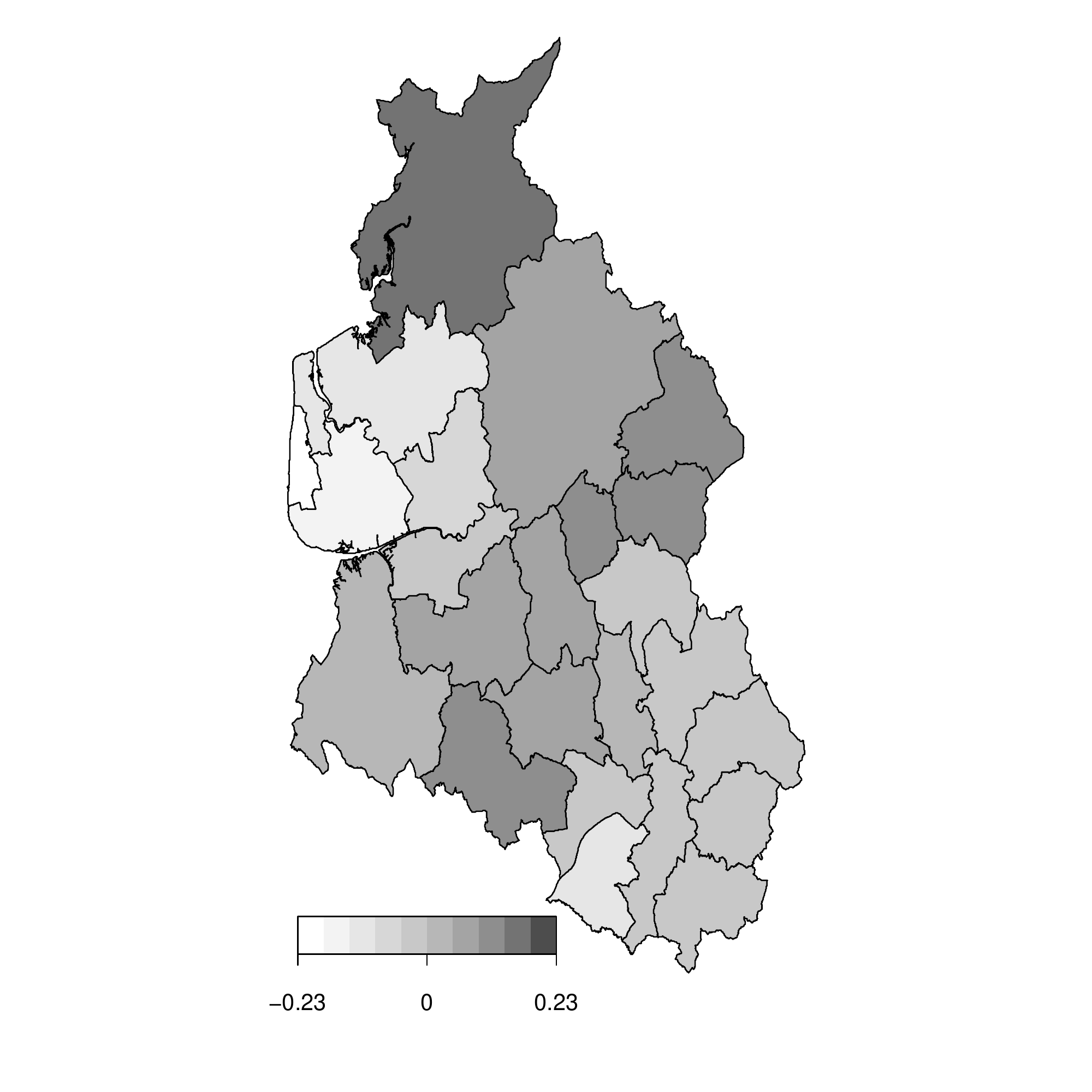} }
	\caption{Leukemia survival data. PO model with ICAR frailties. (a) Cox-Snell plot. (b) Survival curves with $95\%$ credible interval bands for female patients with \code{wbc}=38.59 and \code{tpi}=0.3398 at different ages. (c) Map for the posterior mean frailties; larger frailties mean higher mortality rate overall.}
	\label{leukemia:snell:surv:map}
\end{figure}

The code below is used to generate the Cox-Snell plots with $10$ posterior residuals (Figure~\ref{leukemia:snell:surv:map}, panel a). 
\begin{CodeInput}
R> set.seed(1)
R> cox.snell.survregbayes(res1, ncurves = 10)
\end{CodeInput}

The code below is used to generate survival curves for female patients with \code{wbc}=38.59 and \code{tpi}=0.3398 at different ages (Figure~\ref{leukemia:snell:surv:map}, panel b). 
\begin{CodeInput}
R> tgrid <- seq(0.1, 5000, length.out = 300);
R> xpred <- data.frame(age = c(49, 65, 74), sex = c(0, 0, 0),
+    wbc = c(38.59, 38.59, 38.59), tpi = c(0.3398, 0.3398, 0.3398),
+    row.names = c("age=49", "age=65", "age=74"))
R> plot(res1, xnewdata = xpred, tgrid = tgrid, cex = 2)
\end{CodeInput}

The code below is used to generate the map of posterior means of frailties for each district (Figure~\ref{leukemia:snell:surv:map}, panel c). Note that the posterior median of frailties can be extracted similarly by replacing \code{mean} below with \code{median} in the \code{apply} function.
\begin{CodeInput}
R> frail0 <- apply(res1$v, 1, mean)
R> frail <- frail0[as.integer(names(nwengland))] 
R> values <- cbind(as.integer(names(nwengland)), frail) 
R> op <- par(no.readonly = TRUE)
R> par(mar = c(3, 0, 0, 0))
R> plotmap(nwengland, x = values, col = (gray.colors(10, 0.3, 1))[10:1], 
+    pos = "bottomleft", width = 0.5, height = 0.04)
\end{CodeInput}

\subsubsection{PO model with GRF frailties}
Note that all coordinates are distinct, so we have $m=1043$ and $n_i=1$ in terms of our notation. To use \code{frailtyprior} to specify the prior, we need to create an \code{ID} variable consisting of $1043$ distinct values. The powered exponential correlation function with $\nu=1$ is used. To specify the number of knots and blocks for the FSA of $\bR$, we consider $K=100$ and $B=1043$. The code below is used to fit a PO model with GRF frailties under above settings. The running time is a bit under three hours.  
\begin{CodeChunk}
\begin{CodeInput}
R> set.seed(1)
R> mcmc <- list(nburn = 5000, nsave = 2000, nskip = 4, ndisplay = 1000) 
R> prior <- list(maxL = 15, nu = 1, nknots = 100, nblock = 1043) 
R> d$ID <- 1:nrow(d) 
R> locations <- cbind(d$xcoord, d$ycoord);
R> ptm <- proc.time()
R> res2 <- survregbayes(formula = Surv(time, cens) ~ age + sex + wbc + tpi + 
+    frailtyprior("grf", ID), data = d, survmodel = "PO",
+    dist = "loglogistic", mcmc = mcmc, prior = prior, 
+    Coordinates = locations)
R> proc.time() - ptm 
\end{CodeInput}
\begin{CodeOutput}
     user    system   elapsed 
10079.006    97.039 10176.650 
\end{CodeOutput}
\begin{CodeInput}
R> (sfit2 <- summary(res2)) 
\end{CodeInput}
\begin{CodeOutput}
Posterior inference of regression coefficients
(Adaptive M-H acceptance rate: 0.2726):
      Mean        Median      Std. Dev.   95%CI-Low   95%CI-Upp 
age   0.0526668   0.0527180   0.0034351   0.0460261   0.0596917
sex   0.1310119   0.1318825   0.1069728  -0.0748948   0.3457847
wbc   0.0060590   0.0060293   0.0008156   0.0044876   0.0077388
tpi   0.0606026   0.0609221   0.0158076   0.0300292   0.0918792

Posterior inference of frailty variance
          Mean    Median  Std. Dev.  95%CI-Low  95%CI-Upp
variance  0.06179  0.05290  0.03261    0.02376    0.14086  

Posterior inference of correlation function range phi
        Mean    Median  Std. Dev.  95%CI-Low  95%CI-Upp
range  19.138  17.245   7.305      8.701     35.094   

Log pseudo marginal likelihood: LPML=-5923.402
Deviance Information Criterion: DIC=11845.78
Watanabe-Akaike information criterion: WAIC=11846.78
Number of subjects: n=1043
\end{CodeOutput}
\end{CodeChunk}

\begin{figure}
	\centering
	{ \includegraphics[width = .8\textwidth]{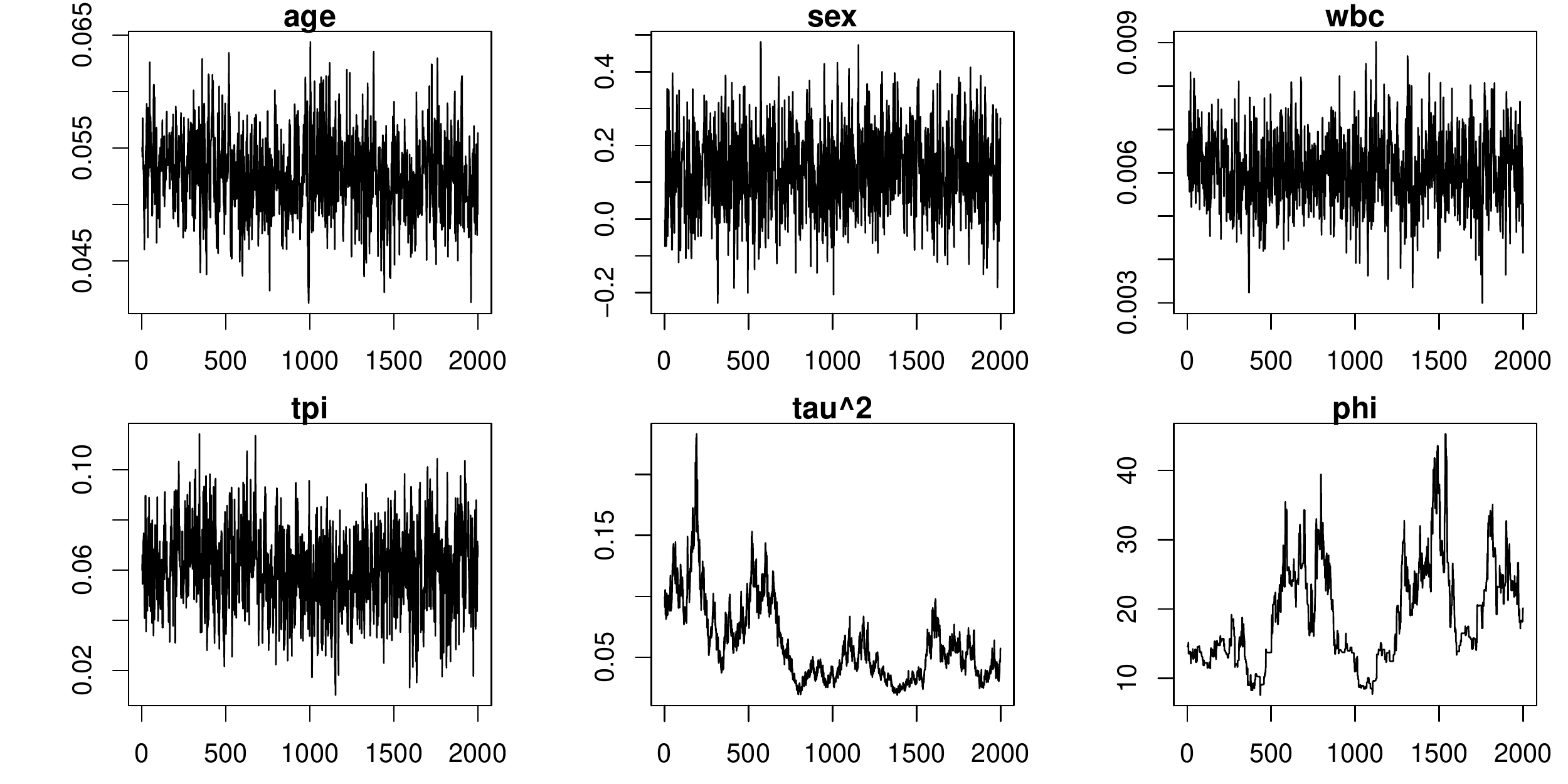} }
	\caption{Leukemia survival data. Trace plots for $\bbeta$, $\tau^2$ and $\alpha$ under the PO model with GRF frailties.}
	\label{leukemia:GRF:trace}
\end{figure}

\begin{figure}
	\centering
	\subfigure[]{ \includegraphics[width=0.3\textwidth]{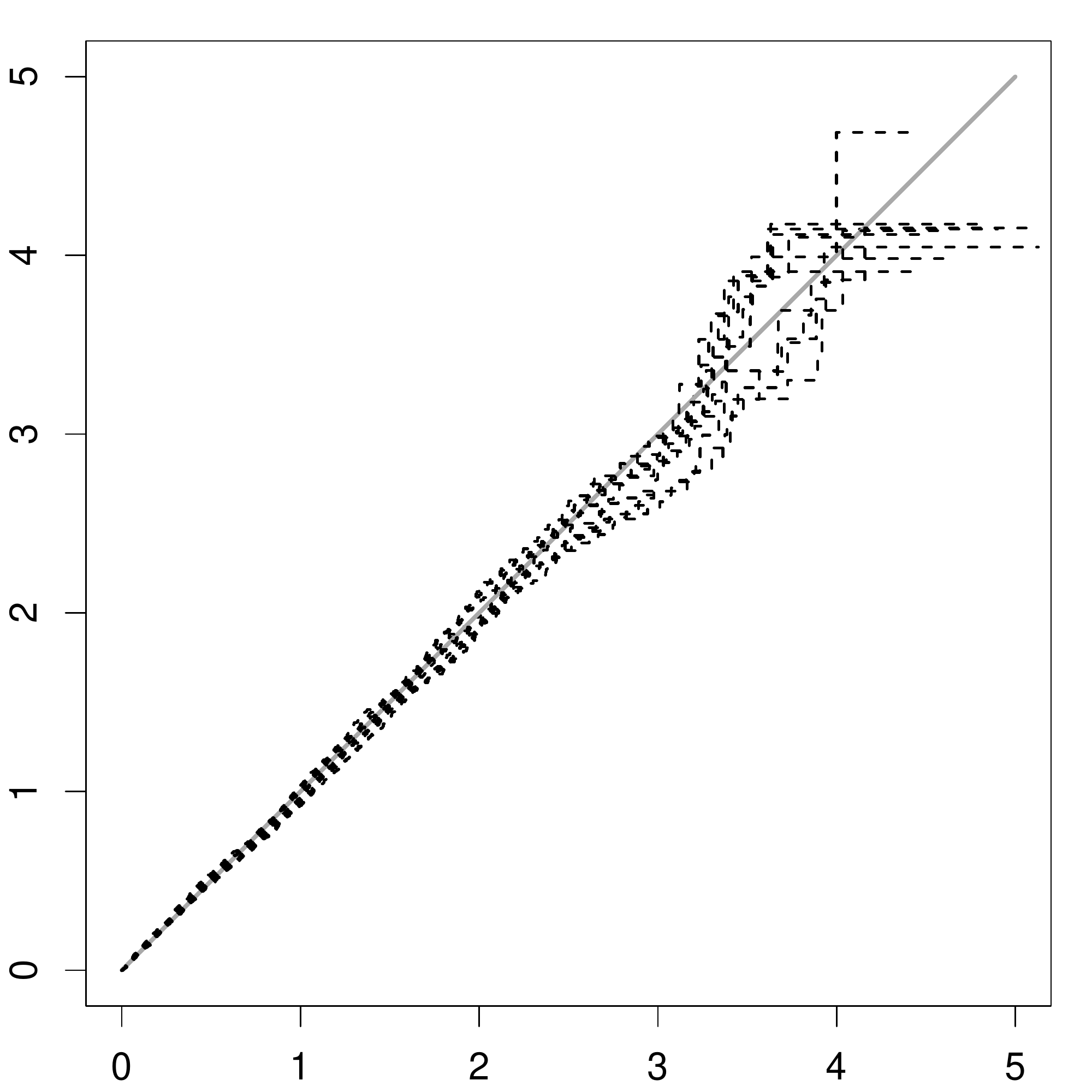} }
	\subfigure[]{ \includegraphics[width=0.3\textwidth]{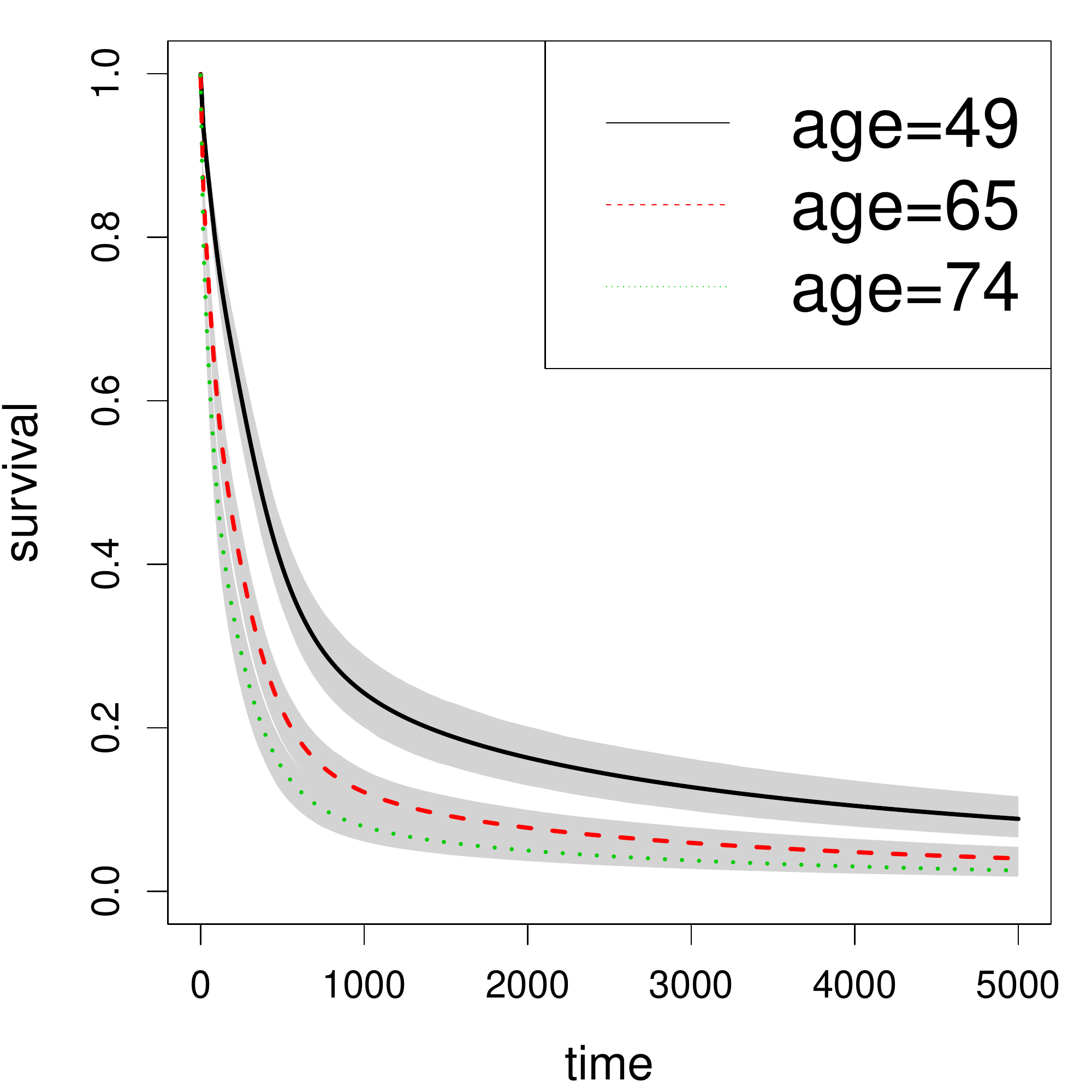} }
	\subfigure[]{ \includegraphics[width=0.3\textwidth]{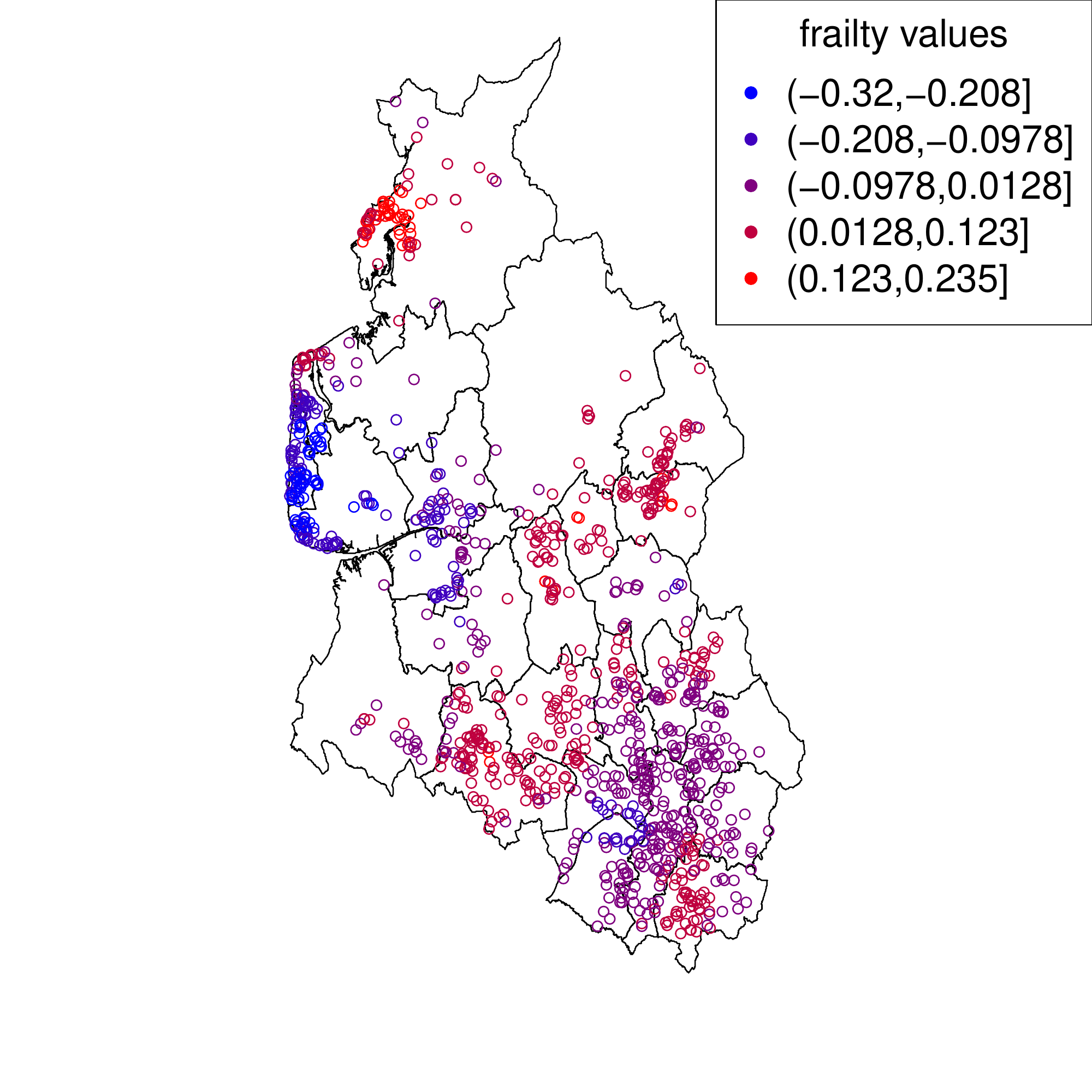} }
	\caption{Leukemia survival data. PO model with GRF frailties. (a) Cox-Snell plot. (b) Survival curves with $95\%$ credible interval bands for female patients with \code{wbc}=38.59 and \code{tpi}=0.3398 at different ages. (c) Map for the posterior mean frailties; larger frailties mean higher mortality rate overall.}
	\label{leukemia:GRF:snell:surv:map}
\end{figure}

The trace plots for $\bbeta$, $\tau^2$ and $\phi$ (Figure~\ref{leukemia:GRF:trace}), Cox-Snell residuals and survival curves (Figure~\ref{leukemia:GRF:snell:surv:map}) can be obtained using the same code used for the PO model with ICAR frailties. The code below is used to generate the map of posterior means of frailties for each location (Figure~\ref{leukemia:GRF:snell:surv:map}). 
\begin{CodeInput}
R> frail <- round(apply(res2$v, 1, mean), 3)
R> nclust <- 5 
R> frail.cluster <- cut(frail, breaks = nclust) 
R> frail.names <- names(table(frail.cluster))
R> rbPal <- colorRampPalette(c('blue', 'red'))
R> frail.colors <- rbPal(nclust)[as.numeric(frail.cluster)]
R> par(mar = c(3, 0, 0, 0))
R> plot(nwengland)
R> points(cbind(d$xcoord,d$ycoord), col = frail.colors)
R> legend("topright", title = "frailty values", legend = frail.names,
+    col = rbPal(nclust), pch = 20, cex = 1.7)
\end{CodeInput}
Note that the mixing for $\tau^2$ and $\phi$ is very poor. This may be partly due to the fact we are updating large dimensional ($m=1,043$) correlated frailties individually using Metropolis-Hastings. From the simulation studies in \cite{Zhou.Hanson2017}, we see that the GRF frailty models perform very well for georeferenced data with replicates at each location. For this dataset, one could create georeferenced data with replicates as follows: group the $1043$ locations into, say $150$, clusters with cluster centroid as the new locations, and assume one shared frailty on each cluster. 

\subsection{Variable selection}\label{sec:selection}
Let $\bx=(x_1,\ldots,x_p)^\top$ denote the $p$-vector of covariates in general. The most direct approach is to multiply $\beta_\ell$ by a latent Bernoulli variable $\gamma_\ell$ for $\ell=1,\ldots,p$, where $\gamma_\ell=1$ indicates the presence of covariate $x_\ell$ in the model, and then assume an appropriate prior on $(\bbeta, \bgamma)$, where $\bgamma=(\gamma_1, \ldots, \gamma_p)^\top$. Following  \cite{Kuo.Mallick1998} and \cite{Hanson.etal2014}, we consider below independent priors 
\begin{equation*} \label{eq:semi:gprior}
	\gamma_1, \ldots, \gamma_p\overset{iid}{\sim} \textrm{Bern}(0.5) \text{ and } \bbeta \sim N_p(\mathbf{0}, g n (\bX^\top\bX)^{-1}),  
\end{equation*}
where $\bX$ is the usual design matrix, but with mean-centered covariates, i.e., 
$\mathbf{1}_n^\top \bX = \mathbf{0}_p^\top$, and $g$ is chosen by picking a number $M$ such that a random $e^{\bx^\top\bbeta}$ is less than $M$ with probability $q$, i.e., approximately $g = \left[ {\log M}/{\Phi^{-1}(q)} \right]^2/{p}$. The function \texttt{survregbayes} sets $M=10$ and $q=0.9$ as the defaults. For other choices, one can specify $M$ and $q$ via \code{prior$M} and \code{prior$q}, respectively.  The MCMC procedure is described in \cite{Zhou.Hanson2017}. 

To perform variable selection for the leukemia survival data, we simply need to add the argument \code{selection=TRUE} to the function \code{survregbayes}. A part of the output from \code{summary} is also shown. The model with \code{age}, \code{wbc} and \code{tpi} has the highest proportion (89.8\%), and thus can be served as the final model. 
\begin{CodeChunk}
\begin{CodeInput}
R> set.seed(1)
R> res3 <- survregbayes(formula = Surv(time, cens) ~ age + sex + wbc + tpi +
+    frailtyprior("car", district), data = d, survmodel = "PO",
+    dist = "loglogistic", mcmc = mcmc, prior = prior, Proximity = E,
+    selection = TRUE) 
R> (sfit3 <- summary(res3))
\end{CodeInput}
\begin{CodeOutput}
Variable selection:
       age,wbc,tpi  age,sex,wbc,tpi  age,wbc
prop.  0.8975       0.1010           0.0015 
\end{CodeOutput}
\end{CodeChunk}

\subsection{Parametric vs. semiparametric}
Many authors have found parametric models to fit as well or better than competing semiparametric models (\citealp[p. 123]{Cox.Oakes1984}; \citealp{Nardi.Schemper2003}). The semiparametric -- or more accurately richly parametric -- formulation of the AFT, PH and PO models presented here have their baseline survival functions centered at a parametric family $S_{\btheta}(t)$. Note that $\bz_{J-1}=\bzero$ implies $S_0(t)=S_{\btheta}(t)$. Therefore, testing $H_0:\bz_{J-1}=\bzero$ versus $H_1:\bz_{J-1}\neq\bzero$ leads to the comparison of the semiparametric model with the underlying parametric model. Let $BF_{10}$ be the Bayes factor between $H_1$ and $H_0$. \cite{Zhou.etal2017} proposed to estimate $BF_{10}$ by a large-sample approximation to the generalized Savage-Dickey density ratio \citep{Verdinelli.Wasserman1995}. Adapting their approach $BF_{10}$ is estimated
\begin{equation*}
\widehat{BF}_{10} = \frac{p(\bzero|\hat{\alpha})}{N_{J-1}(\bzero; \hat{\bfm}, \hat{\bSigma})},
\end{equation*}
where $p(\bzero|\alpha)=\Gamma(\alpha J)/[J^{\alpha}\Gamma(\alpha) ]^J$ is the prior density of $\bz_{J-1}$ evaluated at $\bz_{J-1}=\bzero$, $\hat{\alpha}$ is the posterior mean of $\alpha$, $N_{p}(\cdot; \bfm, \bSigma)$ denotes a $p$-variable normal density with mean $\bfm$ and covariance $\bSigma$, and $\hat{\bfm}$ and $\hat{\bSigma}$ are posterior mean and covariance of $\bz_{J-1}$. 

The Bayes factor $BF_{10}$ under the semiparametric PO model with ICAR frailties can be obtained using the code below (here the object \code{res1} is obtained in Section~\ref{sec:semi:leukemia}).
\begin{Code}
R> BF.survregbayes(res1)
[1] 82.12799
\end{Code}
The $BF_{10}=82>1$ indicates that the semiparametric model outperforms the loglogistic parametric model. 

The function \texttt{survregbayes} also supports the efficient fitting of parametric frailty models with loglogistic, lognormal or Weibull baseline functions. In parametric models, the prior for $\btheta$ can be set to be relatively vague. Setting $a_0$ at any negative value will force the $\alpha$ to be fixed at the value specified in the argument \code{state}. For example, setting \code{prior <- list(a0 = -1)} and \code{state = list(alpha = 1)} will fix $\alpha=1$ throughout the MCMC; setting \code{prior = list(a0 = -1)} and \code{state = list(alpha = Inf)} will fit a parametric model. The following code fits a parametric loglogistic PO model with ICAR frailties to the leukemia survival data. The LPML is -5950, much worse than the value under the semiparametric PO model. 
\begin{CodeChunk}
\begin{CodeInput}
R> set.seed(1)
R> prior <- list(maxL = 15, a0 = -1, thete0 = rep(0, 2), V0 = diag(1e10, 2)) 
R> state <- list(alpha = Inf) 
R> ptm <- proc.time()
R> res11 <- survregbayes(formula = Surv(time, cens) ~ age + sex + wbc + tpi +
+    frailtyprior("car", district), data = d, survmodel = "PO",
+    dist = "loglogistic", mcmc = mcmc, prior = prior, state = state,
+    Proximity = E, InitParamMCMC = FALSE)
R> proc.time() - ptm 
\end{CodeInput}
\begin{CodeOutput}
  user  system elapsed 
 25.037   0.115  25.239 
\end{CodeOutput}
\begin{CodeInput}
R> (sfit11 <- summary(res11))  
\end{CodeInput}
\begin{CodeOutput}
Proportional Odds model:
Call:
survregbayes(formula = Surv(time, cens) ~ age + sex + wbc + tpi + 
    frailtyprior("car", district), data = d, survmodel = "PO", 
    dist = "loglogistic", mcmc = mcmc, prior = prior, state = state, 
    Proximity = E, InitParamMCMC = FALSE)

Posterior inference of regression coefficients
(Adaptive M-H acceptance rate: 0.2844):
      Mean        Median      Std. Dev.   95%CI-Low   95%CI-Upp 
age   0.0504253   0.0504362   0.0033318   0.0439945   0.0568477
sex   0.1187297   0.1134544   0.1109109  -0.0912841   0.3374972
wbc   0.0062192   0.0062147   0.0007395   0.0048068   0.0076600
tpi   0.0602207   0.0603376   0.0156038   0.0299010   0.0915584

Posterior inference of conditional CAR frailty variance
          Mean     Median   Std. Dev.  95%CI-Low  95%CI-Upp
variance  0.078627  0.055078  0.082164   0.002005   0.305202 

Log pseudo marginal likelihood: LPML=-5949.919
Deviance Information Criterion: DIC=11899.46
Watanabe-Akaike information criterion: WAIC=11899.84
Number of subjects: n=1043
\end{CodeOutput}
\end{CodeChunk}

\subsection{Left-truncation and time-dependent covariates}
The survival time $t_{ij}$ is left-truncated at $u_{ij}\geq 0$ if $u_{ij}$ is the time when the $ij$th subject is first observed. Left-truncation often occurs when age is used as the time scale. Given the observed left-truncated data $\{ (u_{ij}, a_{ij}, b_{ij}, \bx_{ij}, \bfs_i)\}$, where $a_{ij}\geq u_{ij}$, the likelihood function in Equation~\ref{like} becomes 
\begin{equation*}\label{like-truncation}
L(\bw_J, \btheta, \bbeta, \bv)=\prod_{i=1}^m \prod_{j=1}^{n_i} \left[S_{\bx_{ij}}(a_{ij})-S_{\bx_{ij}}(b_{ij})\right]^{I\{a_{ij}<b_{ij}\}} f_{\bx_{ij}}(a_{ij})^{I\{a_{ij}=b_{ij}\}}/S_{\bx_{ij}}(u_{ij}).
\end{equation*}
Note that the left censored data under left-truncation are of the form $(u_{ij}, b_{ij})$. 
Allowing for left-truncation allows the semiparametric AFT, PH and PO models to be easily extended to handle time-dependent covariates. Following \cite{Kneib2006} and \cite{Hanson.etal2009}, assume the covariate vector $\bx_{ij}(t)$ is a step function that changes at $o_{ij}$ ordered times $t_{ij,1}<\ldots<t_{ij,o_{ij}}\leq a_{ij}$,  i.e., 
\[ \bx_{ij}(t) = \sum_{k=1}^{o_{ij}} \bx_{ij,k} I(t_{ij,k}\leq t < t_{ij,k+1}),
\] where $t_{ij,1}=u_{ij}$ and $t_{ij,o_{ij}+1}=\infty$.  Assuming one of PH, PO, or AFT holds conditionally on each interval, the survival function for the $ij$th individual at time $a_{ij}$ is
\begin{align*} 
P(t_{ij}>a_{ij}) 
& = P(t_{ij}>a_{ij}|t_{ij}>t_{ij,o_{ij}}) \prod_{k=1}^{o_{ij}-1}P(t_{ij}>t_{ij,k+1}|t_{ij}>t_{ij,k}) \\
& = \frac{S_{\bx_{ij,o_{ij}}}(a_{ij})}{S_{\bx_{ij,o_{ij}}}(t_{ij,o_{ij}})}
\prod_{k=1}^{o_{ij}-1} \frac{S_{\bx_{ij,k}}(t_{ij,k+1})}{S_{\bx_{ij,k}}(t_{ij,k})}.
\end{align*} 
Thus one can replace the observation $(u_{ij}, a_{ij}, b_{ij}, \bx_{ij}(t), \bfs_i)$ by a set of new $o_{ij}$ observations $(t_{ij,1}, t_{ij,2}, \infty, \bx_{ij,1}, \bfs_i)$, $(t_{ij,2}, t_{ij,3}, \infty, \bx_{ij,2}, \bfs_i)$, $\ldots$, $(t_{ij,o_{ij}}, a_{ij}, b_{ij}, \bx_{ij,o_{ij}}, \bfs_i)$. This way we get a new left-truncated data set of size $\sum_{i=1}^{m}\sum_{j=1}^{n_i} o_{ij}$.  Then the likelihood function becomes
\begin{align*}
L(\bw_J, \btheta, \bbeta, \bv) = & \prod_{i=1}^m \prod_{j=1}^{n_i} \bigg\{ \left[S_{\bx_{ij,o_{ij}}}(a_{ij})-S_{\bx_{ij,o_{ij}}}(b_{ij})\right]^{I\{a_{ij}<b_{ij}\}}  f_{\bx_{ij,o_{ij}}}(a_{ij})^{I\{a_{ij}=b_{ij}\}}/S_{\bx_{ij,o_{ij}}}(t_{ij,o_{ij}})  \\
&\times \prod_{k=1}^{o_{ij}-1} \frac{S_{\bx_{ij,k}}(t_{ij,k+1})}{S_{\bx_{ij,k}}(t_{ij,k})}\bigg\} . 
\end{align*}
Note that the derivations above still hold for time-dependent covariates without left-truncation (i.e., $u_{ij}=0$ for all $i$ and $j$).

\subsubsection{PBC data}
We use the primary biliary cirrhosis (PBC) dataset (available in the package \texttt{survival} as \code{pbc}) as an example to show how to incorporate time-dependent covariates in the function \texttt{survregbayes}. Although this is not a spatial dataset, spatial frailties can be added similarly as in Section~\ref{sec:semi:leukemia}. The following code is copied from \cite{Therneau.etal2017} to create the data frame with time-dependent covariates. 
\begin{CodeChunk}
\begin{CodeInput}
R> temp <- subset(pbc, id <= 312, select = c(id:sex, stage)) # baseline data
R> pbc2 <- tmerge(temp, temp, id = id, endpt = event(time, status))
R> pbc2 <- tmerge(pbc2, pbcseq, id = id, ascites = tdc(day, ascites),
+    bili = tdc(day, bili), albumin = tdc(day, albumin),
+    protime = tdc(day, protime), alk.phos = tdc(day, alk.phos))
R> pbc2 <- pbc2[,c("id", "tstart", "tstop", "endpt", "bili", "protime")] 
R> head(pbc2) 
\end{CodeInput}
\begin{CodeOutput}
  id tstart tstop endpt bili protime
1  1      0   192     0 14.5    12.2
2  1    192   400     2 21.3    11.2
3  2      0   182     0  1.1    10.6
4  2    182   365     0  0.8    11.0
5  2    365   768     0  1.0    11.6
6  2    768  1790     0  1.9    10.6
\end{CodeOutput}
\end{CodeChunk}

We can fit the Bayesian PH model with TBP baseline as follows. The output for regression coefficients is partial. 
\begin{CodeChunk}
\begin{CodeInput}
R> set.seed(1)
R> mcmc <- list(nburn = 5000, nsave = 2000, nskip = 4, ndisplay = 1000) 
R> ptm <- proc.time()
R> fit1 <- survregbayes(Surv(tstart, tstop, endpt == 2) ~ log(bili) + 
+    log(protime), data = pbc2, survmodel = "PH", dist = "loglogistic", 
+    mcmc = mcmc, subject.num = id) 
R> proc.time() - ptm
\end{CodeInput}
\begin{CodeOutput}
   user  system elapsed 
227.626   0.434 228.243 
\end{CodeOutput}
\begin{CodeInput}
R> summary(fit1)
\end{CodeInput}
\begin{CodeOutput}
Proportional hazards model:
Call:
survregbayes(formula = Surv(tstart, tstop, endpt == 2) ~ log(bili) + 
    log(protime), data = pbc2, survmodel = "PH", dist = "loglogistic", 
    mcmc = mcmc, subject.num = id)

Posterior inference of regression coefficients
(Adaptive M-H acceptance rate: 0.2135):
              Mean     Median   Std. Dev.  95%CI-Low  95%CI-Upp
log(bili)     1.29937  1.30058  0.09452    1.11354    1.48584  
log(protime)  4.18500  4.20421  0.37052    3.43850    4.84161  

Log pseudo marginal likelihood: LPML=-1018.001
Deviance Information Criterion: DIC=2032.754
Watanabe-Akaike information criterion: WAIC=2035.861
Number of subjects: n=1807
\end{CodeOutput}
\end{CodeChunk}
Equivalently, one can also run the following code to obtain the same analysis. The argument \code{truncation_time} is used to specify the start time point for each time interval, i.e., \code{tstart}. The end time point \code{tstop} together with \code{endpt} are formulated as interval censored data using \code{type = "interval2"} of \code{Surv}. This format is more general than the former one, as one can easily incorporate interval censored data.
\begin{CodeInput}
R> pbc2$tleft <- pbc2$tstop; pbc2$tright <- pbc2$tstop;
R> pbc2$tright[which(pbc2$endpt! = 2)] <- NA;
R> fit11 <- survregbayes(Surv(tleft, tright, type = "interval2") ~ log(bili) +
+    log(protime), data = pbc2, survmodel = "PH", dist = "loglogistic", 
+    mcmc = mcmc, truncation_time = tstart, subject.num = id);
\end{CodeInput}

\section{GAFT frailty models}
\subsection{The model}
The generalized accelerated failure time (GAFT) frailty model \citep{Zhou.etal2017} generalizes the AFT model in Equation~\ref{aft} to allow the baseline survival function $S_0(t)$ to depend on certain covariates, say a $q$-dimensional vector $\bz_{ij}$ which is usually a subset of $\bx_{ij}$. Specifically, the GAFT frailty model is given by 
\begin{equation*}\label{eq:GAFT}
S_{\bx_{ij}}(t)=S_{0,\bz_{ij}}\left(e^{-\bx_{ij}^\top{\bbeta} - v_i}t\right), 
\end{equation*}
or equivalently, 
\begin{equation*}\label{eq:GAFT2}
y_{ij} = \log(t_{ij}) = \tilde{\bx}_{ij}^\top{\tilde{\bbeta}} + v_i + \epsilon_{ij},
\end{equation*}
where $\tilde{\bx}_{ij}=(1,\bx_{ij}^\top)^\top$ includes an intercept,  $\tilde{\bbeta}=(\beta_0,\bbeta^\top)^\top$ is a vector of corresponding coefficients, $\epsilon_{ij}$ is a heteroscedastic error term independent of $v_i$, and $P(e^{\beta_0+\epsilon_{ij}}>t|\bz_{ij})=S_{0,\bz_{ij}}(t)$. Note the regression coefficients $\bbeta$ here are defined differently with those in Equation~\ref{aft}. Here we assume 
$$\epsilon_{ij} | G_{\bz_{ij}} \stackrel{ind.}{\sim} G_{\bz_{ij}},$$
where $G_{\bz}$ is a probability measure defined on $\mathR$ for every $\bz\in\calX$; this defines a model for the entire collection of probability measures $\calG_\calX = \{G_{\bz}: \bz\in\calX\}$ so that each element is allowed to smoothly change with the covariates $\bz$. The \texttt{frailtyGAFT} function considers the following prior distributions:
\begin{equation*}\label{eq:gaft:priors}
\begin{aligned}
\tilde{\bbeta} & \sim N_{p+1}(\bfm_0, \bS_0)\\
G_{\bz} |\alpha, \sigma^2 &\sim \LDTFP_L(\alpha, \sigma^2), ~\alpha \sim {\Gamma} (a_0, b_0), ~\sigma^{-2} \sim \Gamma (a_\sigma, b_\sigma),\\
(v_1, \ldots, v_m)^\top|\tau &\sim \ICAR(\tau^2), ~\tau^{-2}\sim \Gamma(a_\tau, b_\tau) , \text{ or}\\
(v_1, \ldots, v_m)^\top|\tau,\phi &\sim \GRF(\tau^2, \phi), ~\tau^{-2}\sim \Gamma(a_\tau, b_\tau),  ~\phi\sim \Gamma(a_\phi, b_\phi), \text{ or}\\
(v_1, \ldots, v_m)^\top|\tau &\sim \IID(\tau^2), ~\tau^{-2}\sim \Gamma(a_\tau, b_\tau)
\end{aligned}
\end{equation*}
where $\LDTFP_L$ refers to the linear dependent tailfree process prior (LDTFP) prior as described in \citep{Zhou.etal2017}. The function argument \texttt{prior} allows users to specify these prior parameters in a list with elements defined as follows:
\begin{center}
	\centering
	\begin{tabular}{ c | c c ccccccccc} 
		\hline
		element & \texttt{maxL} & \texttt{m0} & \texttt{S0}  & \texttt{a0} & \texttt{b0} & \texttt{siga0} & \texttt{sigb0} & \texttt{taua0} & \texttt{taub0} & \texttt{phia0} & \texttt{phib0}\\
		symbol & $L$ & $\bfm_0$ & $\bS_0$ & $a_0$ & $b_0$ & $a_\sigma$ & $b_\sigma$ & $a_\tau$ & $b_\tau$  & $a_\phi$ & $b_\phi$\\
		\hline
	\end{tabular}
\end{center}

The LDTFP prior considered in \cite{Zhou.etal2017} is centered at a normal distribution $\Phi_\sigma$ with mean $0$ and variance $\sigma^2$, that is, $E(G_{\bz}) = \Phi_\sigma$ for every $\bz\in\calX$. Define the function $k_\sigma(x)=\lceil 2^L \Phi_\sigma(x) \rceil$, where $\lceil x \rceil$ is the ceiling function, the smallest integer greater than or equal to $x$. Further define probability $p_{\bz}(k)$ for $k=1, \ldots, 2^L$ as
\[ p_{\bz}(k) = \prod_{l=1}^L Y_{l,\lceil k2^{l-L}  \rceil}(\bz),\] 
where $ Y_{j+1,2k-1}(\bz) = \left(1+\exp\{-\tilde{\bz}^\top \bgamma_{j,k} \}\right)^{-1}$ and  $Y_{j+1,2k}(\bz)=1-Y_{j+1,2k-1}(\bz)$ for $j=0,\ldots,L-1$, $k=1,\ldots,2^{j}$, where $\tilde{\bz}=(1,\bz^\top)^\top$ includes an intercept, and $\bgamma_{j,k}=(\gamma_{j,k,0}, \ldots, \gamma_{j,k,q})^\top$ is a vector of coefficients. Note there are $2^L-1$ regression coefficient vectors $\bgamma=\{\bgamma_{j,k}\}$, e.g., for $L=3$, $\bgamma=\{\bgamma_{0,1},\bgamma_{1,1},\bgamma_{1,2},\bgamma_{2,1},\bgamma_{2,2},\bgamma_{2,3},\bgamma_{2,4}\}$. For a fixed integer $L>0$, the random density associated with $\LDTFP_L(\alpha, \sigma^2)$ is defined as 
\begin{equation*}\label{ldtfp:f0}
f_{\bz}(e) =  2^L \phi_\sigma(e) p_{\bz}\{k_\sigma(e)\}, ~\bgamma_{j,k} \overset{ind.}{\sim} N_{q+1}\left(\mathbf{0}, \frac{2n}{\alpha (j+1)^2} (\bZ^\top\bZ)^{-1}\right)
\end{equation*}
with cdf 
\begin{equation}\label{ldtfp:S0}
G_{\bz}(e) =  p_{\bz}\{k_\sigma(e)\}\left\{2^L\Phi_\sigma(e)-k_\sigma(e)\right\} + \sum_{k=1}^{k_\sigma(e)}p_{\bz}(k), 
\end{equation}
where $\bZ$ is the $n\times (q+1)$ design matrix with mean-centered covariates $\tilde{\bz}_{ij}$s.  Furthermore, the LDTFP is specified by setting $\bgamma_{0,1}\equiv \bzero$, such that for every $\bz\in\calX$, $G_{\bz}$ is almost surely a median-zero probability measure. 

The function \texttt{frailtyGAFT} sets the following hyperparameters as defaults: $\bfm_0=\bzero$, $\bS_0=10^{5}\bI_{p+1}$, $a_0=b_0=1$, $a_\tau=b_\tau=1$, and $a_\sigma=2+\hat{\sigma}^4_0/(100\hat{v}_0)$, $b_\sigma=\hat{\sigma}_0^2(a_\sigma-1)$, where $\hat{\sigma}_0^2$ and $\hat{v}_0$ are the estimates of $\sigma^2$ and its asymptotic variance from fitting the parametric lognormal AFT model, respectively. Note here we assume a somewhat informative prior on $\sigma^2$ so that its mean is $\hat{\sigma}_0^2$ and variance is $100\hat{v}_0$. For the GRF prior, we again set $a_\phi=2$ and $b_{\phi}=(a_{\phi}-1)/\phi_0$ so that the prior of $\phi$ has mode at ${\phi}_0$ and the prior mean of $1/\phi$ is $1/\phi_0$ with infinite variance. Here $\phi_0$ satisfies $\rho(\bfs',\bfs''; \phi_0)=0.001$, where $\|\bfs'-\bfs''\|=\max_{ij}\|\bfs_i-\bfs_j\|$.  Note by default \code{frailtyGAFT} standardizes each covariate by subtracting the sample mean and dividing the sample standard deviation. Therefore, the user-specified hyperparameters should be based on the model with scaled covariates unless the argument \code{scale.designX = FALSE} is added. 

\subsection{Bayesian hypothesis testing}\label{subsec:BF}
The GAFT frailty model includes the following as important special cases: an AFT frailty model with nonparametric baseline where $G_{\bz}=G_{\bz'}$ for all $\bz=\bz'$ and parametric baseline model $G_{\bz}=\Phi_\sigma$ for all $\bz\in\calX$. Hypothesis tests can be constructed based on the LDTFP coefficients $\{\bgamma_{l,k}: k=1,\ldots,2^l, l=1,\ldots,L-1\}$, where $\bgamma_{l,k}=(\gamma_{l,k,0}, \ldots, \gamma_{l,k,q})^\top$. Let $\bgamma_{l,k,-j}$ denote the subvector of $\bgamma_{l,k}$ without element $\gamma_{l,k,j}$ for $j=0, \ldots, q$. Set $\bUpsilon_j=(\gamma_{l,k,j}, k=1,\ldots,2^l, l=1,\ldots,L-1)^\top$, $\bUpsilon_{-j}=(\bgamma_{l,k,-j}^\top, k=1,\ldots,2^l, l=1,\ldots,L-1)^\top$ and $\bUpsilon=(\bgamma_{l,k}^\top, k=1,\ldots,2^l, l=1,\ldots,L-1)^\top$. Testing the hypotheses $H_0:\bUpsilon_{-0}=\bzero$ and $H_0:\bUpsilon=\bzero$ leads to global comparisons of the proposed model with the above two special cases respectively. Similarly, we may also test the null hypothesis $H_0: \bUpsilon_j=\bzero$ for the $j$th covariate effect of $\bz$ on the baseline survival, $j=1, \ldots, q$.  

Suppose we wish to test $H_0:\bUpsilon_j=\bzero$ versus $H_1:\bUpsilon_j\ne\bzero$, for fixed $j\in\{1,\ldots,q\}$. Following \cite{Zhou.etal2017}, the Bayes factor between hypotheses $H_1$ and $H_0$ can be approximated by
\begin{equation*}\label{eq:BFhat}
\hat{BF}_{10} =  \frac{\ds\prod_{l=1}^{L-1}\prod_{k=1}^{2^l} N\left(0\bigg|0, \frac{2n}{\hat{\alpha} (l+1)^2} (\bZ^\top\bZ)_{jj}^{-1}\right)}{N_{2^{L}-2}(\bUpsilon_j=\bzero; \hat{\bfm}_{j}, \hat{\bS}_{j})},
\end{equation*}
where  $N_{p}(\cdot; \bfm,\bS)$ denotes a $p$-variate normal density with mean $\bfm$ and covariance matrix $\bS$, and $\hat{\bfm}_{j}$ and $\hat{\bS}_{j}$ are the sample mean and covariance for $\bUpsilon_j$.  

\subsection{Leukemia survival data}
\begin{figure}
	\centering\subfigure[]{ \includegraphics[width=0.3\textwidth]{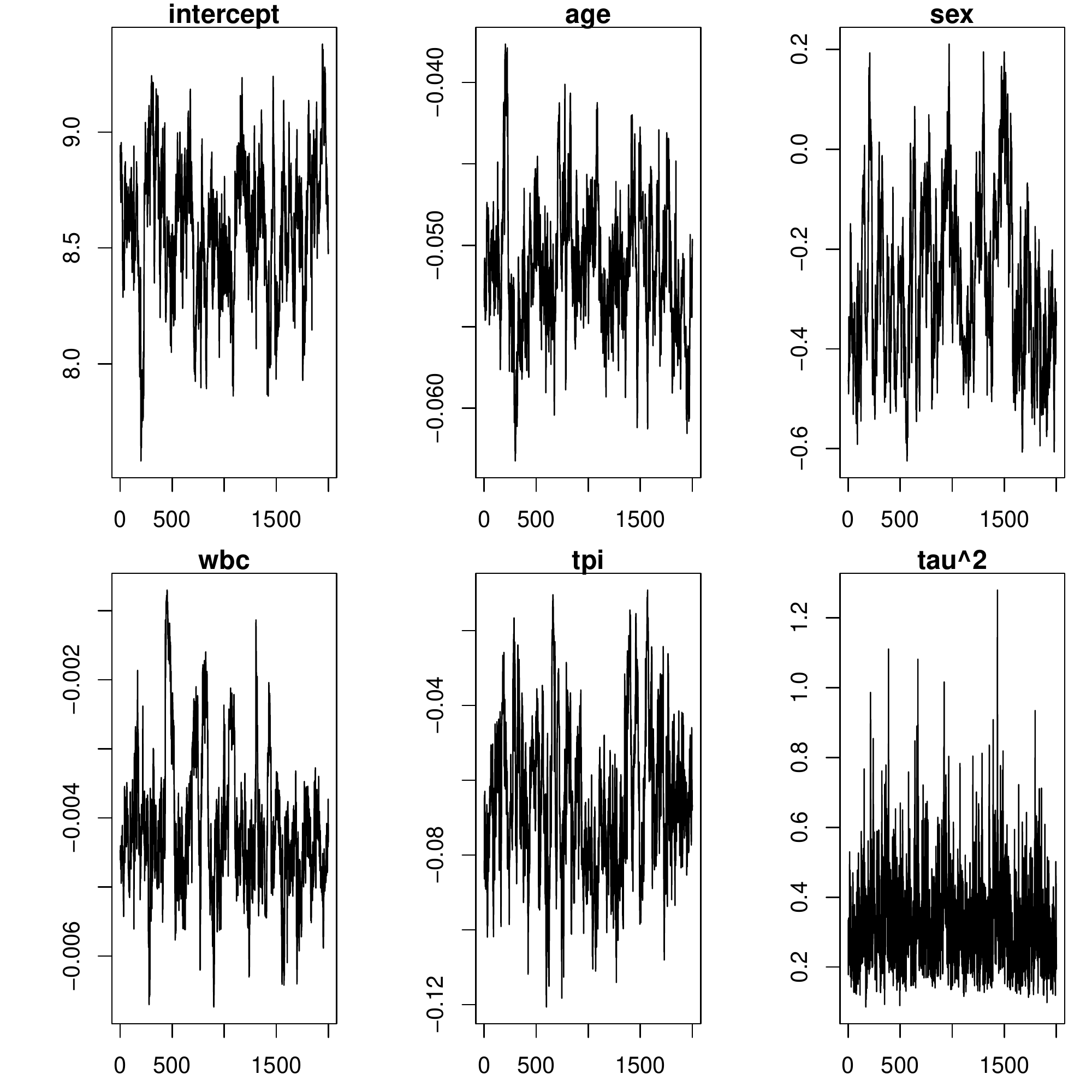} }
	\subfigure[]{ \includegraphics[width=0.3\textwidth]{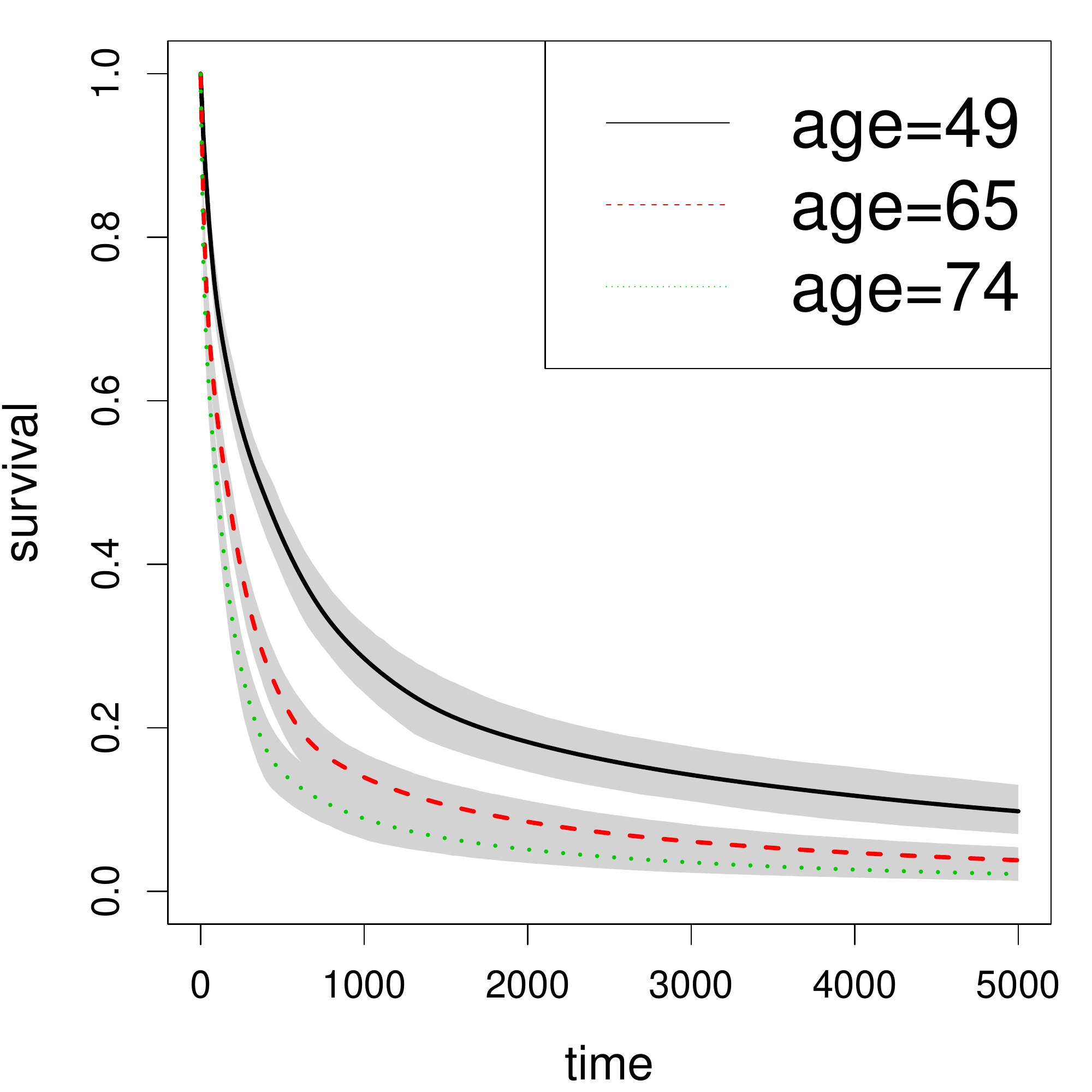} }
	\subfigure[]{ \includegraphics[width=0.3\textwidth]{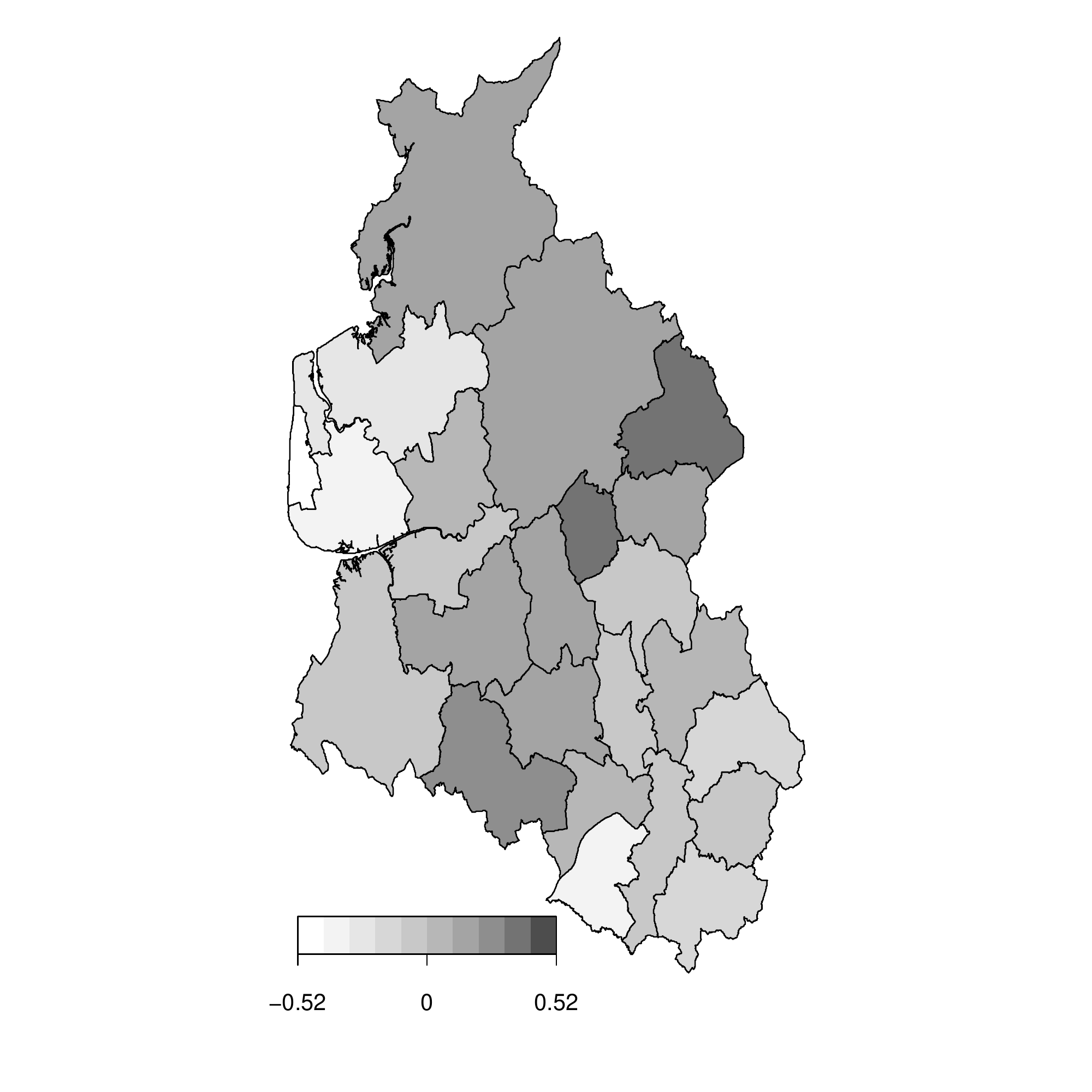} }
	\caption{Leukemia survival data. GAFT model with ICAR frailties. (a) Trace plots for $\bbeta$, $\tau^2$ and $\alpha$. (b) Survival curves with $95\%$ credible interval bands for female patients with \code{wbc}=38.59 and \code{tpi}=0.3398 at different ages. (c) Map for the negative posterior mean frailties; larger values mean higher mortality rate overall.}
	\label{leukemia:GAFTcar:surv:map}
\end{figure}

The code below is used to fit the GAFT model with ICAR frailties for the leukemia survival data. As suggested by \cite{Zhou.etal2017}, the gamma prior $\Gamma(a_0=5, b_0=1)$ is used for $\alpha$. We include all four covariates in modeling the baseline survival function. 
\begin{CodeChunk}
\begin{CodeInput}
R> set.seed(1)
R> mcmc <- list(nburn = 5000, nsave = 2000, nskip = 4, ndisplay = 1000) 
R> prior <- list(maxL = 4, a0 = 5, b0 = 1) 
R> ptm <- proc.time()
R> res1 <- frailtyGAFT(formula = Surv(time, cens) ~ age + sex + wbc + tpi +
+    baseline(age, sex, wbc, tpi) + frailtyprior("car", district),
+    data = d, mcmc = mcmc, prior = prior, Proximity = E) 

R> (sfit1 <- summary(res1)) ## Output below is partial
\end{CodeInput}
\begin{CodeOutput}
Generalized accelerated failure time frailty model:
Call:
frailtyGAFT(formula = Surv(time, cens) ~ age + sex + wbc + tpi + 
    baseline(age, sex, wbc, tpi) + frailtyprior("car", district), 
    data = d, mcmc = mcmc, prior = prior, Proximity = E)

Posterior inference of regression coefficients
            Mean       Median     Std. Dev.  95%HPD-Low  95%HPD-Upp
intercept   8.589761   8.607783   0.288535   7.982265    9.140489 
age        -0.051342  -0.051508   0.003987  -0.058561   -0.041985 
sex        -0.267978  -0.288883   0.164310  -0.533180    0.064909 
wbc        -0.004161  -0.004322   0.001001  -0.005931   -0.001864 
tpi        -0.065335  -0.067061   0.019601  -0.099992   -0.023739 

Bayes factors for LDTFP covariate effects:
intercept        age        sex        wbc        tpi    overall  normality  
 220.2500    16.2494     1.1579    28.1776     0.4842    11.2454  1787.3269  

Log pseudo marginal likelihood: LPML=-5937.304
Number of subjects:=1043
\end{CodeOutput}
\begin{CodeInput}
R> proc.time() - ptm
\end{CodeInput}
\begin{CodeOutput}
   user  system elapsed 
444.393   2.433 454.270 
\end{CodeOutput}
\end{CodeChunk}
The Bayes factors for testing \code{age} and \code{wbc} effects on LDTFP are 16 and 28, respectively, indicating that the baseline survival function under the AFT model depends on \code{age} and \code{wbc}, and thus GAFT should be considered.  The trace plots, survival curves and frailty map (Figure~\ref{leukemia:GAFTcar:surv:map}) can be obtained using the code similarly as in Section~\ref{sec:semi:leukemia}. The only difference for plotting survival curves is that we need to specify the baseline covariates by including the argument \code{xtfnewdata = xpred} into the \code{plot} function. Note that the mixing for covariate effects is okay but not great due to the non-smoothness of Polya trees. In this case, we need to run a longer chain with much higher thinning as suggested in \cite{Zhou.etal2017}.

\section{Survival models via spatial copulas} 
In environmental studies, survival times (e.g. time to water pollution) often present a strong spatial dependence after adjusting for available risk factors, making frailty models extremely difficult to fit because of the strong posterior dependency among frailties. The spatial copula approach \citep{Bardossy2006} offers an appealing way to describe spatial dependence among survival times separately from their univariate distributions, thus leads to more efficient posterior sampling algorithms. In addition, the regression coefficients have population-level interpretations under copula models. However, the copula approach can be very slow in the presence of high censoring rate due to the imputation of centered survival times. 

Currently the package only supports spatial copula models for georeferenced (without replication, i.e., $n_i=1$), right-censored spatial data. Suppose subjects are observed at $n$ distinct spatial locations $\bfs_1,\dots,\bfs_n$. Let $t_{i}$ be a random event time associated with the subject at $\bfs_i$ and $\bx_{i}$ be a related $p$-dimensional vector of covariates, $i=1, \ldots, n$. For right-censored data, we only observe $t_i^o$ and a censoring indicator $\delta_i$ for each subject, where $\delta_i$ equals 1 if $t_i^o=t_i$ and equals 0 if $t_i$ is censored at $t_i^o$. Therefore, the observed data will be $\mathcal{D}=\{(t_{i}^o,\delta_{i},\bx_i, \bfs_i); i=1,\ldots, n\}$. Note although the models below are developed for spatial survival data, non-spatial data are also accommodated. 

In the context of survival models, the idea of spatial copula approach is to first assume that the survival time $t_i$ at location $\bfs_i$ marginally follows a model $S_{\bx_i}(t)$, then model the joint distribution of $(t_1, \ldots, t_n)^\top$ as 
\begin{equation*}\label{CopulaModel}
	P(t_1\leq a_1, \ldots, t_n\leq a_n) = C(F_{\bx_1}(a_1), \ldots, F_{\bx_n}(a_n)),
\end{equation*}
where $F_{\bx_i}(t)=1-S_{\bx_i}(t)$ is the cumulative distribution function and the function $C$ is an $n$-copula used to capture spatial dependence.  

The current package assumes a spatial version of the Gaussian copula \citep{Li2010}, defined as
\begin{equation}\label{GaussianCopula:cdf}
C(u_1, \ldots, u_n) = \Phi_n\left( \Phi^{-1}\{u_1\}, \ldots, \Phi^{-1}\{u_n\}; \bR\right),
\end{equation}
where $\Phi_n(\cdot,\ldots,\cdot; \bR)$ denotes the distribution function of $N_n(\mathbf{0}, \bR)$. To allow for a nugget effect, we consider  $\bR[i,j]=\theta_1\rho(\bfs_i, \bfs_j;\theta_2)+(1-\theta_1)I(\bfs_i=\bfs_j)$, where $\rho(\bfs_i, \bfs_j;\theta_2)=\exp\{-\theta_2\|\bfs_i-\bfs_j\|\}$. Here $\theta_1\in[0,1]$, also known as a ``partial sill'' in \cite{Waller.Gotway2004}, is a scale parameter measuring a local maximum correlation, and $\theta_2$ controls the spatial decay over distance. Note that all the diagonal elements of $\bR$ are ones, so it is also a correlation matrix. Under the above spatial Gaussian copula, the likelihood function based on upon the complete data $\{(t_i, \bx_i, \bfs_i), i=1, \ldots, n\}$ is
\begin{equation*}\label{Likelihood}
\calL = |\bR|^{-1/2}\exp\left\{-\frac{1}{2}\bz^\top(\bR^{-1}-\bI_n)\bz\right\}\prod_{i=1}^n f_{\bx_i}(t_i),
\end{equation*}
where $z_i=\Phi^{-1}\left\{F_{\bx_i}(t_i)\right\}$ and $f_{\bx_i}(t)$ is the density function corresponding to $S_{\bx_i}(t)$. We next discuss two marginal spatial survival models for $S_{\bx_i}(t)$ that are accommodated in the package. Note that for large $n$, the FSA introduced in Section~\ref{sec:semi:frailty} (with $\epsilon$ replaced by $1-\theta_1$) can be applied.

\subsection{Proportional hazards model via spatial copulas}
Assume that $t_i|\bx_i$ marginally follows the proportional hazards (PH) model with cdf
\begin{equation}\label{spCopulaCoxph}
F_{\bx_i}(t) = 1 - \exp\left\{-\Lambda_0(t)e^{\bx_i^\top\bbeta}\right\}
\end{equation}
and density 
\begin{equation*}
f_{\bx_i}(t) = \exp\left\{-\Lambda_0(t)e^{\bx_i^\top\bbeta}\right\}\lambda_0(t)e^{\bx_i^\top\bbeta},
\end{equation*}
where $\bbeta$ is a $p\times 1$ vector of regression coefficients, $\lambda_0(t)$ is the baseline hazard function and $\Lambda_0(t)=\int_0^t \lambda_0(s)ds$ is the cumulative baseline hazard function. The piecewise exponential model provides a flexible framework to deal with the baseline hazard \citep[e.g.,][]{Walker.Mallick1997}. We partition the time period $\mathbb{R}^+$ into $M$ intervals, say $I_k=(d_{k-1}, d_k], k=1, \ldots, M$, where $d_0=0$ and $d_M=\infty$. Specifically, we set $d_k$ to be the $\frac{k}{M}$th quantile of the empirical distribution of the observed survival times for $k=1,\ldots,M-1$. The baseline hazard is then assumed to be constant within each interval, i.e.,
\begin{equation*}
\lambda_0(t) = \sum_{k=1}^M h_k I\{t\in I_k \},
\end{equation*}
where $h_k$s are unknown hazard values. Consequently, the cumulative baseline hazard function can be written as
\begin{equation*}
\Lambda_0(t) = \sum_{k=1}^{M(t)} h_k \Delta_k(t),
\end{equation*}
where $M(t)=\min\{k:d_k \geq t\}$ and $\Delta_{k}(t) = \min\{d_k, t\} - d_{k-1}$.  After incorporating spatial dependence via the copula in Equation~\ref{GaussianCopula:cdf}, the \texttt{spCopulaCoxph} function considers the following prior distributions:
\begin{equation*}\label{eq:CopulaCoxph:priors}
\begin{aligned}
\bbeta &\sim N_p(\bbeta_0, \bS_0),\\
h_k|h &\overset{iid}{\sim}\mathrm{\Gamma}(r_0h, r_0), k=1, \ldots, M, \\
(\theta_1, \theta_2) &\sim \mathrm{Beta}(\theta_{1a}, \theta_{1b}) \times \mathrm{\Gamma}(\theta_{2a}, \theta_{2b})
\end{aligned}
\end{equation*}

The \texttt{spCopulaCoxph} function sets the following default hyperparameter values: $M=10$, $r_0=1$, $h=\hat{h}$, $\bbeta_0=\bzero$, $\bS_0=10^5\bI_p$, $\btheta_0=(\theta_{1a}, \theta_{1b}, \theta_{2a}, \theta_{2b})'=(1,1,1,1)$, where $\hat{h}$ is the maximum likelihood estimate of the rate parameter from fitting an exponential PH model. A function \texttt{indeptCoxph} is also provided to fit the non-spatial standard PH model with above baseline and prior settings. The function argument \texttt{prior} allows users to specify these prior parameters in a list with elements defined as follows:
\begin{center}
	\centering
	\begin{tabular}{ c | c c cccc} 
		\hline
		element & \texttt{M} & \texttt{r0} & \texttt{h0}  & \texttt{beta0} & \texttt{S0} & \texttt{theta0} \\
		symbol & $M$ & $r_0$ & $h$ & $\bbeta_0$ & $\bS_0$ & $\btheta_0$ \\
		\hline
	\end{tabular}
\end{center}

\subsection{Bayesian nonparametric survival model via spatial copulas}
We assume that $y_i=\log t_i$ given $\bx_i$ marginally follows a LDDPM model \citep{DeIorio.etal2009} with cdf,
\begin{equation}\label{eq:LDDPM}
F_{\bx_i}(t) = \int \Phi\left(\frac{\log t-\bx_i^\top\bbeta}{\sigma}\right) d G\{\bbeta, \sigma^2\},
\end{equation}
where $\Phi(\cdot)$ is the cdf of the standard normal, and $G$ follows the Dirichlet Process (DP) prior. This Bayesian nonparametric model treats the conditional distribution $F_{\bx}$ as a function-valued parameter and allows its variance, skewness, modality and other features to flexibly vary with the $\bx$ covariates. After incorporating spatial dependence via the copula in Equation~\ref{GaussianCopula:cdf}, the function \texttt{spCopulaDDP} assumes the following prior distributions:
\begin{equation*}\label{eq:CopulaDDP:priors}
\begin{aligned}
G &= \sum_{k=1}^N w_k \delta_{(\bbeta_k, \sigma_k^{2})}, ~w_k = V_k \prod_{j=0}^{k-1} (1-V_j), ~ V_0=0, V_N=1\\
V_k &\overset{iid}{\sim}\mathrm{Beta}(1, \alpha), k=1, \ldots, N, ~ \alpha \sim \Gamma(a_0, b_0) \\
\bbeta_k|\bmu & \overset{iid}{\sim} N_p(\bmu, \bSigma), k=1, \ldots, N, ~\bmu \sim N_p(\bfm_0, \bS_0)\\
\sigma^{-2}_k |\bSigma & \overset{iid}{\sim} \Gamma(\nu_{a}, \nu_{b}), k=1, \ldots, N, ~ \bSigma^{-1} \sim W_p\left( (\kappa_0\bSigma_0)^{-1}, \kappa_0 \right)\\
(\theta_1, \theta_2) &\sim \mathrm{Beta}(\theta_{1a}, \theta_{1b}) \times \mathrm{\Gamma}(\theta_{2a}, \theta_{2b}).
\end{aligned}
\end{equation*}
The following default hyperparameters are considered in \texttt{spCopulaDDP}: $a_0=b_0=2$, $\nu_a=3$, $\nu_b=\hat{\sigma}^2$, $\btheta_0=(\theta_{1a}, \theta_{1b}, \theta_{2a}, \theta_{2b})'=(1,1,1,1)$, $\bfm_0=\hat{\bbeta}$, $\bS_0=\hat{\bSigma}$, $\bSigma_0=30\hat{\bSigma}$, and $\kappa_0=7$, where $\hat{\bbeta}$ and $\hat{\sigma}^2$ are the maximum likelihood estimates of $\bbeta$ and $\sigma^2$ from fitting the log-normal accelerated failure time model $\log(t_i)=\bx_i^\top\bbeta + \sigma\epsilon_i, \epsilon_i\sim N(0,1)$, and $\hat{\bSigma}$ is the asymptotic covariance estimate for $\hat{\bbeta}$. A function \texttt{anovaDDP} is also provided to fit the non-spatial LDDPM model in Equation~\ref{eq:LDDPM} with above prior settings. The function argument \texttt{prior} allows users to specify these prior parameters in a list with elements defined as follows:
\begin{center}
	\centering
	\begin{tabular}{ c | c c ccccccc} 
		\hline
		element & \texttt{N} & \texttt{a0} & \texttt{b0}  &\texttt{m0} & \texttt{S0} & \texttt{k0} & \texttt{Sig0} & \texttt{theta0} \\
		symbol & $N$ & $a_0$ & $b_0$ & $\bfm_0$ & $\bS_0$& $\kappa_0$ & $\bSigma_0$ & $\btheta_0$ \\
		\hline
	\end{tabular}
\end{center}

\subsection{Leukemia survival data}
\subsubsection{PH model with spatial copula}
The following code is used to fit the piecewise exponential PH model in  Equation~\ref{spCopulaCoxph} with the Gaussian spatial copula in Equation~\ref{GaussianCopula:cdf} using $M=20$ and default priors. We consider $K=100$ and $B=1043$ for the number of knots and blocks in the FSA of $\bR$. The total running time is 15445 seconds. 
\begin{CodeChunk}
\begin{CodeInput}
R> set.seed(1)
R> mcmc <- list(nburn = 5000, nsave = 2000, nskip = 4, ndisplay = 1000);
R> prior <- list(M = 20, nknots = 100, nblock = 1043);
R> ptm <- proc.time()
R> res1 <- spCopulaCoxph(formula = Surv(time, cens) ~ age + sex + wbc + tpi, 
+    data = d, mcmc = mcmc, prior = prior,
+    Coordinates = cbind(d$xcoord, d$ycoord));
R> proc.time() - ptm
\end{CodeInput}
\begin{CodeOutput}
     user    system   elapsed 
15262.274   177.716 15444.913 
\end{CodeOutput}
\begin{CodeInput}
R> (sfit1 <- summary(res1)) 
\end{CodeInput}
\begin{CodeOutput}
Spatial Copula Cox PH model with piecewise constant baseline hazards
Call:
spCopulaCoxph(formula = Surv(time, cens) ~ age + sex + wbc + 
     tpi, data = d, mcmc = mcmc, prior = prior, Coordinates = cbind(d$xcoord, 
     d$ycoord))

Posterior inference of regression coefficients
(Adaptive M-H acceptance rate: 0.2501):
      Mean        Median      Std. Dev.   95%CI-Low   95%CI-Upp 
age   0.0277864   0.0278065   0.0019297   0.0240332   0.0315580
sex   0.0522938   0.0527421   0.0588919  -0.0625843   0.1662136
wbc   0.0027808   0.0027899   0.0003767   0.0020071   0.0034546
tpi   0.0257918   0.0257969   0.0081385   0.0087972   0.0411955

Posterior inference of spatial sill and range parameters
(Adaptive M-H acceptance rate: 0.2112):
       Mean     Median   Std. Dev.  95%CI-Low  95%CI-Upp
sill   0.23051  0.23352  0.05587    0.10222    0.32903  
range  0.41801  0.34165  0.34272    0.03715    1.31802  

Log pseudo marginal likelihood: LPML=-5929.357
Number of subjects: n=1043
\end{CodeOutput}
\end{CodeChunk}

Note that the higher the value of $z_i=\Phi^{-1}\left\{ F_{\bx_i}(t_i) \right\}$ is, the longer the survival time $t_i$ (i.e., lower mortality rate) would be. The posterior sample of $z_i$s is saved in \code{res1$Zpred}. The trace plots, survival curves, and the map of the posterior mean of $z_i$ values can be obtained using the code similarly as in Section~\ref{sec:semi:leukemia}.  

\subsubsection{LDDPM model with spatial copula}
The following code is used to fit the LDDPM model in Equation~\ref{eq:LDDPM} with the Gaussian spatial copula in Equation~\ref{GaussianCopula:cdf} using $N=10$ and default priors. For the FSA, $K=100$ and $B=1043$ are used. The total running time is 20056 seconds. Note there is no \code{summary} output as before, as we are fitting a nonparametric model. The trace plots, survival curves, and map of $z_i$s can be obtained using the same code used for the PH copula model.
\begin{CodeChunk}
\begin{CodeInput}
R> set.seed(1)
R> mcmc <- list(nburn = 5000, nsave = 2000, nskip = 4, ndisplay = 1000) 
R> prior <- list(N = 10, nknots = 100, nblock = 1043) 
R> ptm <- proc.time()
R> res1 <- spCopulaDDP(formula = Surv(time, cens) ~ age + sex + wbc + tpi, 
+    data = d, mcmc = mcmc, prior = prior,
+    Coordinates = cbind(d$xcoord, d$ycoord)) 
R> proc.time() - ptm
\end{CodeInput}
\begin{CodeOutput}
     user    system   elapsed 
19876.947   178.595 20056.744 
\end{CodeOutput}
\begin{CodeInput}
R> sum(log(res1$cpo)); ## LPML 
\end{CodeInput}
\begin{CodeOutput}
[1] -5931.5
\end{CodeOutput}
\end{CodeChunk} 

\section{Conclusions}
There is a wealth of \proglang{R} packages for non-spatial survival data, starting with \pkg{survival}, included with all base installs of \proglang{R}.  The \pkg{survival} package fits (discretely) stratified semiparametric PH models to right-censored data with exchangeable gamma frailties, as well as left-truncated data, time-dependent covariates, etc.  Parametric log-logistic, Weibull and log-normal AFT models can also be fit by this package.  From there, there are many packages for various models and types of censoring; a partial review discussing several available \proglang{R} packages is given by \cite{Zhou.Hanson2015}; also see \cite{Zhou.Hanson2017}.  In comparison there are very few \proglang{R} packages for spatially correlated survival data, with the notable exceptions of \pkg{R2BayesX} and \pkg{spatsurv}, both of which focus on PH exclusively.  The \pkg{spBayesSurv} package allows the routine fitting of several popular semiparametric and nonparametric models to spatial survival data.

\pkg{spBayesSurv} can also handle non-spatial survival data using either exchangeable Gaussian or no frailty models. Another unintroduced function is \texttt{survregbayes2} which implements the Polya tree based PH, PO, and AFT models of \cite{Hanson2006} and \cite{Zhao.etal2009} for areally-referenced data.  As pointed out in these papers, MCMC mixing for Polya tree models can be highly problematic when the true baseline survival function is very different from the parametric family that centers the Polya tree; the TBP prior provides much improved MCMC mixing with essentially the same quality of fit as Polya trees.  Another function very recently added function is \texttt{SuperSurvRegBayes}, which provides Bayes factors for testing among PO, PH, and AFT, as well as three other survival models \cite{Zhang.etal2018}.

Future additions to \pkg{spBayesSurv} include spatial copula (both georeferenced and areal) versions of the PH, PO, and AFT models using TBP priors, as well as continuously-stratified proportional hazards and proportional odds models.  An extension of all semiparametric models to additive linear structure, which is already incorporated into \proglang{BayesX}, is also planned.  Finally, computational efficiency can be gained by replacing some of the adaptive MCMC updates with gradient-based updates for the semiparametric models, e.g. the IWLS updates implemented in \proglang{BayesX} for the PH model \citep{Hennerfeind.etal2006}.

\section*{Acknowledgments}
This research is partially funded by Grant R03CA176739 from National Institutes of Health. The authors would like to thank referees for their valuable comments, and all users who have reported bugs and given suggestions.

\bibliographystyle{jss}
\bibliography{bibliography}

\end{document}